\documentclass{ar-1col}

\usepackage{url}
\usepackage[numbers]{natbib}

\jname{Xxxx. Xxx. Xxx. Xxx.}
\jvol{AA}
\jyear{YYYY}
\doi{10.1146/((please add article doi))}

\usepackage{graphicx} 
\usepackage{color}
\usepackage{amsmath}
\usepackage{amsfonts}
\usepackage{bm}
\usepackage{ulem}
\usepackage{placeins}

\renewcommand{\theequation}{\arabic{equation}}

\newcommand{\K}{K$^{+}$}
\newcommand{\Na}{Na$^{+}$}
\newcommand{\Sr}{Sr$^{2+}$}
\newcommand{\Mg}{Mg$^{2+}$}
\newcommand{\Ca}{Ca$^{2+}$}
\newcommand{\Ba}{Ba$^{2+}$}

\begin{document}

\markboth{}{Biomolecular Hydration Mimicry}

\title{Hydration Mimicry by Membrane Ion Channels}

\author{Mangesh I. Chaudhari,$^1$ Juan M. Vanegas,$^1$
L. R. Pratt,$^2$ Ajay Muralidharan,$^2$
and Susan B. Rempe$^1$
\affil{$^1$Computational Biology and Biophysics Department, Sandia National Laboratories, Albuquerque, NM, 87185}
\affil{$^2$Department of Chemical and Biomolecular Engineering, Tulane
University, New Orleans, LA, 70118}
}

\begin{abstract}
Ions transiting biomembranes might  pass readily from water through
ion-specific  membrane proteins if those protein channels provide environments
similar to the aqueous solution hydration environment. Indeed,
bulk aqueous
solution is an important reference condition for the ion permeation
process. 
Assessment of this hydration mimicry view depends on
understanding the hydration structure and free energies of metal ions in
water to provide a comparison for the membrane channel environment.
To refine these considerations, we review local hydration structures of
ions in bulk water, and the molecular quasi-chemical theory that
provides hydration free energies.  In that process, we note some current
views of ion-binding to membrane channels, and suggest new physical
chemical calculations  and experiments that might further clarify the
hydration mimicry view.
\end{abstract}

\begin{keywords}
biomolecular hydration mimicry, membrane ion channels, hydration of metal ions,
hydration free energy, hydration structure,
quasi-chemical theory, density functional theory, statistical thermodynamic theory,
KcsA, MgtE, Ca$_v$Ab 
\end{keywords}
\maketitle

\tableofcontents


\section{BACKGROUND}
Cells control salt concentration differences across boundary membranes by
transporting ions selectively \cite{hille}. Selective ion transport   plays an important role in numerous
physiological functions, including electrical signaling and cell volume
control. Toward that end, proteins ---
either channels or transporters --- provide pathways for ions to
permeate. Blocking such pathways can have detrimental or beneficial
effects. In beneficial cases, drugs that block specific channels hold
promise for treating neurological disorders, autoimmune diseases, and
cancers \cite{Wulff,Zhao,Pardo}. Peptide toxins from several poisonous
animals provide examples of the detrimental
possibilities \cite{Banerjee,Rosenbaum}. Indeed, simple divalent metal ions 
can be potent channel blockers, and both monovalent and
divalent ions permeate selectively. In addition to important roles in
health, cellular mechanisms of ion transport also can guide materials
science by inspiring \cite{Cygan,Freeman}, or being integrated
into \cite{fane,biomembranes:2018}, synthetic membranes for efficient power generation,
water purification, mineral recovery, and separation of small molecules
from mixtures \cite{Fu:2018}.
       
\begin{figure}[tb]
\includegraphics[height=2.8in]{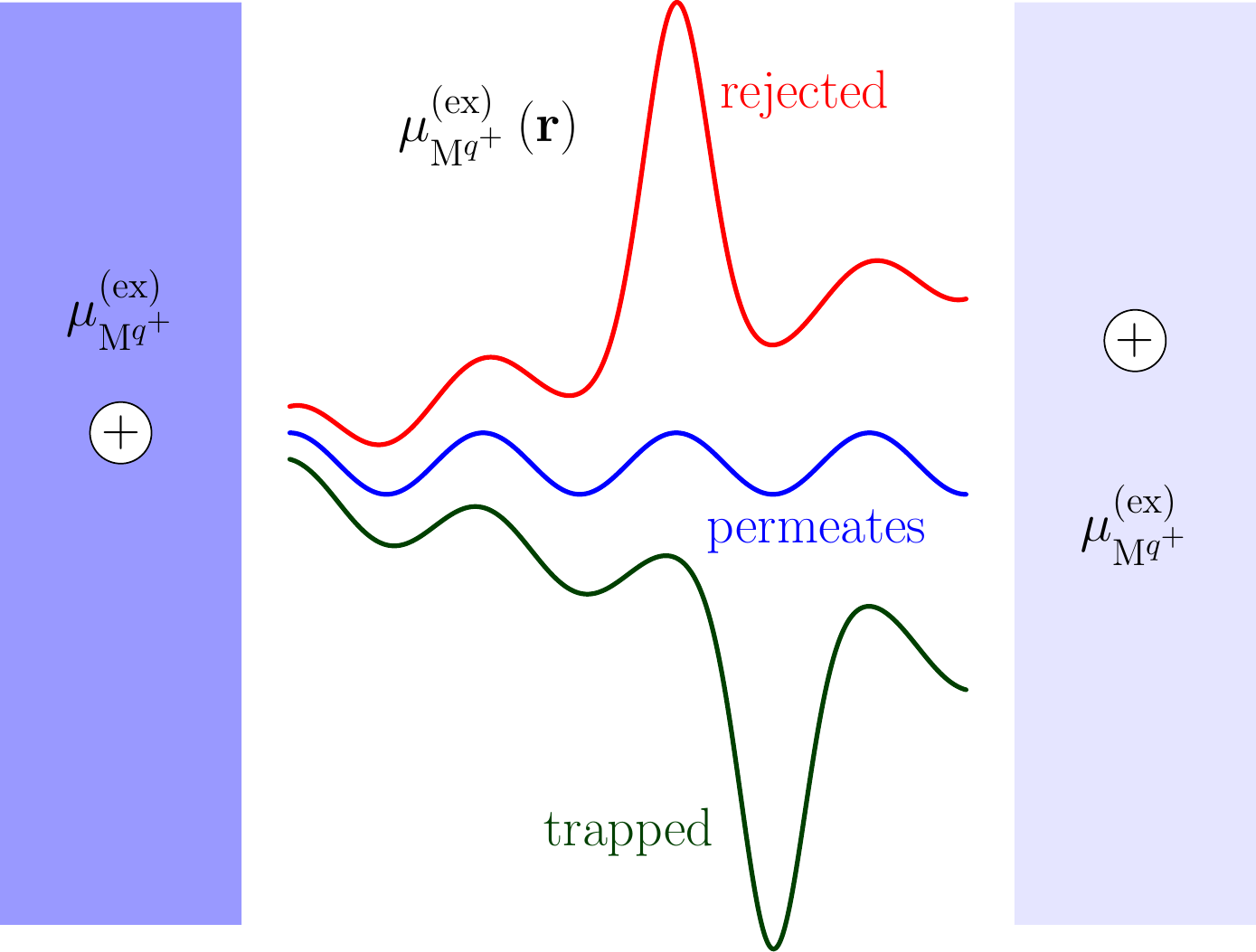}
\caption{Do ion channels mimic the aqueous hydration of
ions that readily permeate? Hypothetical molecular-scale variation 
of the free energies of ion binding, $\mu_{\mathrm{M}^{q+}}
^{(\mathrm{ex})}\left(\mathbf{r}\right)$, illustrate
rejected, permeating, or trapped cases. 
}\label{fig:Hydmim}
\end{figure}

According to a concept called ``hydration mimicry,'' ions might pass easily from water through protein channels
when pore-lining amino acid residues arrange a local environment that
mimics the local ion hydration
environment \cite{Doyle:1998wq,Zhou:2001vo,%
MacKinnon:2003ca,MacKinnon:2004el}.
Here, ``\emph{local}'' structure refers to atoms that interact 
with an ion, making direct contacts. That local structural similarity may  lead to a free
energy for ion binding that approximately equals the free energy for ion
hydration in bulk liquid water, with moderate barriers leading to 
rapid ion permeation (Figure~\ref{fig:Hydmim}). 

Similarly, non-native ions may encounter large free energy barriers,
leading to rejection from the protein pore if a binding site 
offers a poor hydration mimic (Figure~\ref{fig:Hydmim}). 
Alternatively, a non-native ion  
may be trapped by binding  too strongly to a channel (Figure~\ref{fig:Hydmim}), 
and block passage of a native permeant 
ion \cite{asthagiri2006role,Jiang:2000,Jiang:2014,Piasta:2011bu,ye}.

Primitive concepts for design of biomimetic 
transport \cite{rempe2010computational,rempe2016biomimetic} can begin with
focus on the direct contacts of an ion in transit, and the consequences
for the binding free energies. Contacts should be chemically competent
for binding that is satisfactory, but not too strong \cite{pohorille2012water}.
Those binding contacts should be available, and be flexible enough to
accommodate ions in satisfactory binding geometries. Binding geometries may be characterized by properties
like the  number of contacts, or coordination number, and the distance between an ion and its contacts, or cavity size.
Binding geometries also
may be constrained by  properties of the binding sites like covalent bonds, or by interactions with the proximal environment.  Longer-range interactions,
electrostatic and dispersion, are also essential, but perhaps not
decisive in comparing ions with the same charge and similar sizes.  Few approaches can treat all of these features.
Fortunately, recent statistical thermodynamic theory --- quasi-chemical
theory (QCT) --- offers 
a descriptive tool that facilitates analysis of those primitive structural concepts
and their effect on ion binding free energies.  QCT  is a feature
of the presentation that follows.

It is tempting to go further with refined hydration mimicry concepts, 
say,  by noting interesting higher-resolution details of 
ion-protein binding configurations.   But such comparisons 
have been limited by lack of the corresponding higher-resolution 
characterization of the binding of relevant ions to the bulk 
solution, which defines endpoints of ion transfer.  Securing 
the natural 
bulk ion hydration details has been 
a challenge \cite{Rempe:Li,redbook,Varma:2006,Mason:2015kw}. Here, 
we collect and discuss those important results, which have 
been obtained
in recent work.

Theoretical studies \cite{Varma2007,Stevens:2016,chaudhari2018SrBa} that
assess the structure-free energy relationship between ion binding to a
protein,  relative to binding in water, are rarer than direct numerical
simulation of such systems despite an abundance of 
high-resolution cryo-electron microscopy and
crystal structures for channels and
transporters. Here, we analyze the hydration mimicry concept by
comparing local structure exhibited by molecular simulations of ion hydration and crystal structures
of ion binding sites in channels and transporters. We also connect local hydration structure with the free energy
of ion binding to water. Our results emphasize the
importance of coordination number, and of neighborship analyses of molecular
simulations, to characterize local ion hydration structure, thus permitting
comparison with experimental data on local structure for ions in channel and transporter
binding sites. Neighborship analyses also enhance the utility 
of the quasi-chemical free energy theory (QCT) for testing the hydration
mimicry idea. Altogether, our results highlight unexpected relationships
between local solvent structure and transfer free energies for ion
permeation, rejection, and trapping that support an expanded
view of hydration mimicry.  

\subsection{Ions in Water}

Water is the reference environment
for ion transport across cellular membranes.  Transport takes ions from 
one aqueous solution to another, and may bring water molecules along as well \cite{hille,Miller:1982,Latorre:1989,Iwamoto:2011,Armstrong:2015}. Since protein channels and transporters catalyze ion transport across cellular membranes, transport reflects
a balance between ion-water interactions \cite{pohorille2012water} 
and ion-protein interactions \cite{Roux:2005kd}. 
Water molecules
and functional groups from
proteins may interact directly with transiting ions
\cite{Zhou:2001vo,Jiang:2014,Payandeh2013,Tang2014,Lockless:2007ex}, 
providing ligands and defining the 
local ion solvation structure. 
\begin{marginnote}[]
\entry{Bulk aqueous reference}{Bulk hydration studies provide information about the reference environment, \emph{i.e.}, the end points that bracket a transit path 
for ions across membranes.}
\end{marginnote}


        In assessing whether channel and transporter proteins form binding sites 
that mimic the local hydration structures of ions that readily permeate, a 
one ingredient is 
the hydration properties of ions in the reference environment of bulk liquid water.
Ions and water form preferred coordination
structures and those preferred structures may not be attainable in protein environments \cite{Varma2007,Valiyaveetil:2006fy,Varma:2008kl,tilegenova2019structure}.

Studies of ion hydration structure and hydration free energy
benefit from combining experimental and theoretical 
approaches \cite{Varma:2006,Sabo:2013gs,%
Mason:2015,ajay_F}.  In the case of hydration 
structure, experiments can provide
information about average water structure.    Molecular 
simulations, verified against experimental 
data, can resolve
average structure into contributions from individual water molecules. That resolution
identifies local hydration structure, the information sought 
here. The structural predictions can be  scrutinized further by 
using local structures to build predictions of ion hydration free energies, 
also tested for consistency with experimental data in gas and 
liquid phases. In this way, local ion hydration 
structure is reliably connected to ion hydration free energy, providing
insights about the mechanism of ion hydration.  
Further description of this overall approach follows below.

\subsubsection{Hydration structure}
Experiments that probe ion hydration structure include neutron 
and X-ray diffraction techniques that report 
differential scattering cross-sections leading to partial
structure factors \cite{Mason:2015}.  An alternative 
technique, extended X-ray absorption fine 
structure (EXAFS) spectroscopy, also provides data related to partial 
structure factors \cite{Fulton2003,Na_Fulton,K_Fulton}. 
These factors can be determined with good accuracy, and yield
pair-wise radial distribution functions (RDF) between ions and 
water oxygens. 
The RDF establishes information that characterizes 
the structure of solvent around ions of interest. 
For example, integrating
the RDF through an inner-shell
yields  hydration numbers of ions.  

In practice, significant challenges of good spectral resolution 
and momentum cutoffs (for diffraction work) are encountered in 
obtaining 
accurate experimental hydration numbers.  Nevertheless, recent 
developments in both data acquisition and analysis, 
motivated in part by molecular simulation
studies based on \emph{ab initio} approaches and QCT \cite{Rempe:Li}, have 
significantly advanced experimental results \cite{Mason:2015}. 
Even with accurate hydration numbers determined from experiment, 
information 
about local hydration structure may still be lacking.  
\emph{Ab initio} molecular dynamics (AIMD) simulations 
can further differentiate local structural features, \emph{e.g.}
disposition of the \emph{closest} water molecule, from the 
aggregated
hydration structure defined conventionally by the RDFs. 

As with experiments, AIMD 
simulation results are challenged 
by demands of realistic access to suitable ranges of 
space and time \cite{Kirchner:AIMD}.   Nevertheless, 
recent developments in both algorithms and computer power 
have extended that 
ability, leading to improved information on ion hydration structures 
\cite{Li:Laaksonen,Meijer:F,whitfield,Cl:Klein,Lesniewski:2018jd,sharma2018nature,zhou2019importance,karmakar2019ab,martinek2018calcium}. 
Our presentation 
here highlights the correspondence between ion hydration structure 
illuminated by AIMD simulations with available experiment.

To resolve local structure, simulation calculations can  provide 
natural  neighborship analyses of ion hydration structures, 
providing information not readily obtained from experiments.    
Those analyses
distinguish the  water molecules directly contacting an ion from those 
ligands that split time between ion contact and more distant solvent
environments. 
Distinguishing  `non-split shell' and `split-shell' water molecules
helps in
assessing the hydration mimicry idea.  

\begin{marginnote}[]
\entry{Neighborship}{Neighborship analyses describe
the structures of the  ligands closest to an ion, 
and can reveal `\emph{split-shell}' coordination structures. }
\end{marginnote}

While not broadly implemented so 
far
\cite{chaudhari2018SrBa,ajay_F, Rempe:K, Chaudhari:2014wb, Chaudhari:2017gs}, 
neighborship analyses should be more widely used due to 
clear conceptual connection to solvation free energy on the basis of 
molecular quasi-chemical theory. At the same time,  
solvation free
energies computed by QCT test local structural predictions, 
and facilitate the testing of concepts of solvation mechanism  based 
on coordination numbers and other 
primitive concepts described above for design of biomimetic transport.


        \subsubsection{Hydration free energy}
Predictions of ion hydration free energies from molecular simulations face numerous 
challenges. One significant challenge arises because ion-water interactions are 
complicated beyond commonly used molecule-pair interaction 
models.  Those complications reflect solvent polarizability and 
multi-molecule interactions generally, which 
affect interactions in the local solvation 
environment \cite{Varma:2010,rossi}.  
That complexity calls again
for an approach that takes into account electronic degrees of 
freedom \cite{Kirchner:AIMD}.

Another challenge arises with calculations of ion hydration free energies 
with QCT based on treating inner-shell clusters  together with the initial
assumption of small displacements (thus harmonic motions)  of 
water molecules neighboring an ion. Recent work has refined this 
issue, but it deserves further development \cite{ajay_cl}.

Free energy contributions from 
hydration of   
isolated clusters  are computed separately in QCT. 
Properties of   those  clusters are  accessible from experiments, such as high pressure mass spectrometry \cite{Peschke1998}.  The latter 
experiments determine the free energies of equilibrium cluster association 
reactions in gas phase \cite{tissandier1998proton}, and also can be 
compared 
directly with, or even incorporated into, QCT analyses.

Recent work \cite{ajay_F,ajay_cl,chaudhari2017quasi} provided 
examples of coupling between experiment and theory to predict ion
hydration free energies, incorporating gas phase ion clustering
free energies into QCT for halide ions. 
That work accounted for anharmonic vibrational motions 
observed spectroscopically in clusters with several 
coordinating waters \cite{Robertson:2003eu}. 
Further, density functionals used in AIMD simulations  were selected
to reproduce peak positions in experimentally determined 
radial distribution functions (RDF).  
Hydration free energy predictions yielded excellent 
agreement with experimental hydration free energy of 
the neutral LiF pair.  Importantly for the 
present goals, that work followed reliable identification of local hydration 
structure, which led to reliable ion hydration free energies.   Details of the procedure, and applications to a variety of ions, follow below.


    \subsection{Ions in Protein Binding Sites}
Ion channels and transporters are ubiquitous, and
nature has evolved a variety of  membrane proteins 
specialized to the
transport of ions such as \K, \Na, \Ca, and \Mg. The 
history of membrane transport proteins is now seven decades old. Yet, new ion transport proteins, structures,
functions, and mechanisms of both new and \emph{old} transport proteins are
discovered almost daily.  This makes the field of ion channels and transporters
one of the most active in molecular
biology \cite{MacKinnon:2004el,Roux:2005kd,tilegenova2019structure,lockless,
MoraisCabral:2001bp,%
gouaux,%
Beckstein:2003ko,Catterall:2011db,Payandeh:2013gr,Zanni,Cuello,Miller,Mindell,Giraldez,Nimigean,Oster:2019wq}.

Many ion channels  catalyze  rapid transport (10$^6$-10$^8$
ions/s), while simultaneously being highly selective for a specific
ion. Those two properties seem counter-intuitive.  
Experimental results going back to the pioneering work
of Hodgkin and Keynes \cite{Hodgkin:1955tb}, however, indicated that ion
channels that selectively catalyze rapid transport of  ions
have multiple sites where ions
bind \cite{MacKinnon:2004el,Sather2003}. Models based on rate theory
provide an intuitive explanation for the phenomenon of rapid transport
facilitated by multiple binding sites 
\cite{Sather2003,Dang:1998tb}. 
A pre-requisite, 
noted earlier \cite{MacKinnon:2003ca}, 
is that ion binding relative to 
aqueous solution should be weak for rapid 
transport. 

As illustrated in Figure~\ref{fig:Hydmim},  
the free energy wells and barriers for rapid translocation 
should be minimal 
and centered around the free energy for ion hydration in bulk water,
$\mu_{\mathrm{M}^{q+}}^{(\mathrm{ex})}\left(\mathbf{r}\right) \approx
\mu_{\mathrm{M}^{q+}}^{(\mathrm{ex})}$. Another ion may encounter
large energetic barriers
($\mu_{\mathrm{M}^{q+}}^{(\mathrm{ex})}\left(\mathbf{r}\right) \gg
\mu_{\mathrm{M}^{q+}}^{(\mathrm{ex})}$) and, therefore, be rejected.
Alternatively, the channel pore may provide a more favorable environment
compared to the bulk aqueous phase
($\mu_{\mathrm{M}^{q+}}^{(\mathrm{ex})}\left(\mathbf{r}\right) \ll
\mu_{\mathrm{M}^{q+}}^{(\mathrm{ex})}$),  trapping an ion by binding tightly and hence
blocking the channel. 



        \subsubsection{Ion properties}
The hydration mimicry idea originated to 
explain separation between ions of the 
same charge, but different sizes; specifically, for potassium (K)
channels that conduct larger K$^+$ preferentially over smaller Na$^+$. Ions can also be distinguished
by coordination number.   Comparison between the
coordination structure of K$^+$ in a binding site and K$^+$ in 
a neighboring water-filled cavity showed both ions with 8-fold 
coordination \cite{Zhou:2001vo}. Earlier 
analysis of neutron scattering data also proposed 8 as the preferred coordination for 
K$^+$ in 
bulk liquid water \cite{Varma:2006,Ohtomo},  though later work revised
that number \cite{Varma:2006,Neilson:2001,Soper:2006,Soper:2007}.
Altogether, these results supported the initial proposal of hydration mimicry
as a mechanism for rapid and selective ion transport \cite{Doyle:1998wq,Zhou:2001vo,%
MacKinnon:2003ca,MacKinnon:2004el}. 



        \subsubsection{Ligand properties}
Research outside the field of membrane transport proteins has led to a variety of ideas
about which  ligand properties underlie preferential solvation of specific ions, 
but the concepts do not always account for  ion binding preferences demonstrated by protein binding sites.
One prominent concept attributes ligand chemistry as a key factor in
ion binding preferences \cite{Eisenman2}; another attributes the matching of ligand hydration free energy with ion hydration free energy \cite{Collins}.
Recent tests of these concepts on ion binding to binding sites composed of acetate molecules
found no support for the latter concept, called the ``equal affinities'' hypothesis. The same study reported
support for the former concept, the ``ligand field strength'' hypothesis, but only when binding sites included the preferred number of ligands, as determined by free energy analysis.
An important take-home message was that neither concept accounts for the role of the environment on binding site structure.

 In proteins, 
the  matrix surrounding binding sites, or properties of the binding sites 
themselves, may hinder ligand freedom to rearrange upon ion binding.  
Consequently, binding sites may form sub-optimal arrangements for ion
binding, leading to less favorable ion binding.  

Properties of a binding site that naturally take into account constraints from the environment include cavity size, or
distances between ions and ligands, and the number of ligands that coordinate bound ions.
While hydration mimicry is the main topic for testing here, 
cavity size was proposed as an initial explanation for the counter-intuitive
size discrimination of K-channels, focusing attention 
solely on  well-fitting ion-ligand distances \cite{Armstrong:1980,Armstrong:1982}.
In both proposals, local ion solvation structure plays the key role in ion 
selection and, combined with the chemistry of the ligands, compensates 
for ion solvation requirements for rapid permeation \cite{gouaux}. 

These factors of ligand number and cavity size echo 
conclusions from earlier work on ion carriers like the small molecule, valinomycin.
Although valinomycin lacks a transport pathway that would require a specific binding
free energy consistent with permeation (Figure 1), it does 
bind larger K$^+$ selectively over smaller Na$^+$ ions \cite{Dobler}.
Notably, valinomycin binds K$^+$  using fewer ligands than K-channels. 
Constraints on cavity size due to intra-molecular bonds and the
surrounding solvation environment provide a compelling explanation for
selective K$^+$ binding by valinomycin \cite{Varma:2008kl}. In view of
experimental data that characterizes K-channel binding sites as
moderately inflexible when occupied by permeant ions, constrained cavity
size that provides a so-called ``snug fit'' to permeant ions might
account for selective K$^+$ binding by K-channels \cite{McDermott}. A
constrained cavity size, however, may inhibit transport of ions between
well-fitting binding sites \cite{Varma2007}.
        

        

\begin{figure*}[ht]
	\includegraphics[width=\textwidth]{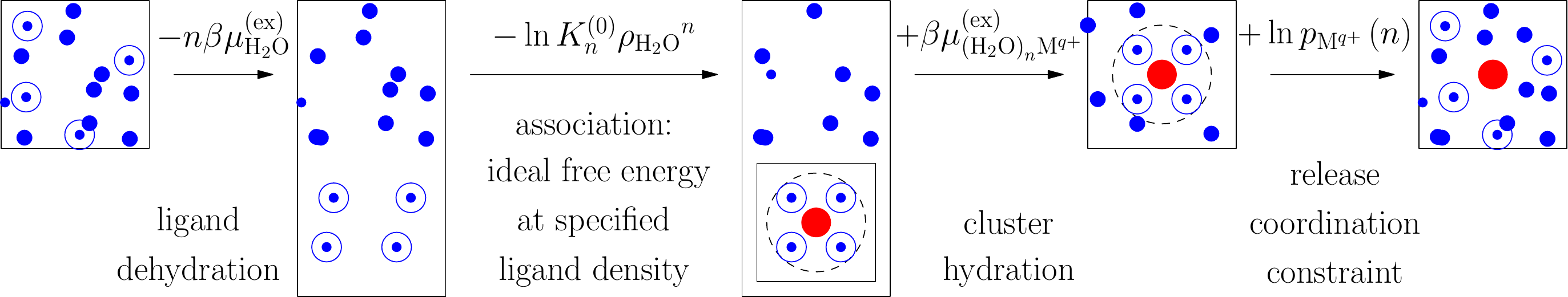}
	\caption{Hydration free energy of solute $\mathrm{M}^{q+}$ 
	(red dot), interpreted on the basis of Eq.~\eqref{eq:1} The central feature of the QCT approach is 
	analysis of the cluster $\mathrm{(H_2O)}_n{\mathrm{M}^{q+}}$. }
	\label{fig:QCT}
\end{figure*}

\subsection{Specific Ions and  Proteins for Hydration Mimicry Analysis}\label{sec:diva}
Ions selected to test the hydration mimicry concept include both monovalent alkali metal ions (Li, Na, K, Rb) and divalent alkaline earth metal ions (Mg, Ca, Sr, Ba). Protein binding sites analyzed here come from
a celebrated bacterial potassium-selective channel (KcsA) \cite{MacKinnon:2004el}, 
the recently discovered bacterial magnesium-selective transporter (MgtE) \cite{Payandeh2013,Takeda2014}, 
and the voltage-gated bacterial calcium-selective channel (Ca$_\mathrm{v}$Ab) \cite{Tang2014,Sather2003}. 

In the results to follow, we report on the size  of the local 
solvation structures and coordination numbers
of the ions examined here. The most probable distances between ions
and ligands, determined  by the first maxima in radial 
distribution functions, and the distances 
reported in crystal structures, measure size.
Cavity sizes should follow these quantities. 
The inner-shell radii, set by the first RDF maxima sets another 
measure of size.  We discuss both of these size measures for 
the ions in liquid water, and
compare them with ion-ligand distances in protein crystal structures 
for insights into hydration mimicry.

Comparisons of ion coordination properties include the number and 
chemistry of ligating atoms.  
Our main interest here for testing the hydration mimicry idea is on number 
of ligating atoms that
directly interact with the ion in stable complexes. Additional
considerations, to be highlighted elsewhere,
include the effect on ion solvation free energies from constraints on local binding site structures that may be imposed
by the surrounding protein environment. 

\subsubsection{Specific ions}
  Among the monovalent ions,  Na$^+$ and K$^+$ are well-known for their 
roles in initiating and terminating
action potentials \cite{hille}.  As analogues to K$^+$ and Na$^+$, Rb$^+$ and Li$^+$ are also 
interesting monovalent ions.  In laboratory settings, Li$^+$ ions 
show contrasting rejection 
behaviors and Rb$^+$ ions  permeate \cite{Lockless:2007ex}. 

Among the divalent ions, \Ca\ and \Mg\ are common 
in biological systems.  
Homeostasis of these ions is tightly controlled by binding 
proteins, channels, exchangers, and pumps or 
transporters \cite{Verkhratsky2014,Carafoli1987,Romani2011}.  
\Sr\ and \Ba\ ions are not biologically relevant, but are well-known
blockers of K$^+$ permeation in K-channels. These ions
are used to understand
the mechanism of K-channel function \cite{Jiang:2000,Jiang:2014,ye,lockless,
Armstrong:1991iq,Elinder:1996cj}. Despite close similarity in size and 
identical +2 charge, \Ba and \Sr exhibit 
different blocking behaviors in 
K-channels \cite{Sugihara:1998vp,Soh:2002ga}.
In particular, \Ba ~blocks bacterial K-channels
like KcsA, but  \Sr does not \cite{Piasta:2011bu}.

The comparison among ions presented here highlights patterns involving 
local solvation structure in
water and channel binding sites among the eight ions selected for study. 
These ions span a range of sizes and charges. 
Ions of similar size include Li$^+$/\Mg
and Na$^+$/\Ca.  Several ions treated here resemble K$^+$ in size ---  
Rb$^+$, \Ba, and \Sr.

\begin{marginnote}[]
\entry{Ion comparisons}{An extended sequence of 
simple metal ions show contrasting rejecting, permeating, 
and trapping behaviors.}
\end{marginnote}

\subsubsection{Specific channels and transporters}\label{sec:proteins}


In potassium-selective channels like the bacterial KcsA protein, crystal structures show dehydrated K$^+$ ions in four
binding sites,
S1-S4 numbered from extracellular to intracellular sides 
\cite{MacKinnon:2003ca}. Structural studies also show dehydrated
Ba$^{2+}$ in the S4 binding site of bacterial K-channels \cite{Jiang:2000,Jiang:2014,ye,lockless}. The environment varies around 
binding sites 
since water borders the ends of the selectivity filter 
(S1, S4) in the open state. Here, we consider an interior binding site (S2) that coordinates permeant ions with carbonyl oxygen atoms from the protein backbone. We also consider the innermost binding site located near bulk liquid water (S4), and composed of the backbone and side chain 
oxygens of four threonines.

Despite conservation of amino acid residues that form K-channel binding 
sites, the function of K-channels can change under certain conditions.  
The bacterial KcsA channel catalyzes rapid passage of K$^+$ across membranes, but rejects the smaller Na$^+$ 
ions by a high ratio of 1000:1 in physiological conditions \cite{hille}.
By simply changing the solution environment  from high  to low
K$^+$ concentration, the 4 conserved K$^+$ binding sites distort so that
the channel switches from a conducting configuration to a non-conducting
configuration \cite{Zhou:2001vo}.
In the unusual absence of K$^+$,  Na$^+$ ions can also distort 
the channel structure and  block  ion 
permeation \cite{MacKinnon:2003ca}.  In related NaK channels, two sites 
are identical to KcsA (S3 and S4);
nevertheless, both K$^+$ and Na$^+$ permeate equally well.  Explanations of 
rapid, selective ion permeation
in potassium channels should  account for these intriguing observations, 
too.


The Ca$_\mathrm{v}\mathrm{Ab}$ crystal structure shows three distinct 
ion binding sites, with the middle
one having the highest affinity. The presence of binding sites 
with different affinities is in good agreement with other experimental 
observations, and a `\emph{stairstep}' 3-site rate theory model can 
explain both
the high selectivity and transport rates observed for Ca$_\mathrm{v}\mathrm{Ab}$ 
channels \cite{Sather2003}. Here,  we select the middle,
highest affinity \Ca~binding site from Ca$_\mathrm{v}\mathrm{Ab}$, which also rejects 
magnesium \cite{Tang2014}.

The Mg-selective channels also 
contain three primary ion binding sites.  While
both \Mg~and \Ca~ions can bind to the M1 binding site,  the M3 binding 
site selectively binds \Mg~ion \cite{Takeda2014}.  For testing
the hydration mimicry idea, we select the \Mg-selective M3 binding site from MgtE.

For the applications here, the free energy for ion binding to a channel 
protein, relative to its hydration free energy, is central to
understanding the thermodynamic driving forces and mechanisms of ion
permeation. Our tool is quasi-chemical theory (QCT), discussed in the 
following. QCT is a
statistical mechanical theory based on close solution contacts treated
at chemical resolution, by {\it  ab initio}
methods \cite{redbook,Asthagiri:2010,%
Rogers:2011}. The exploitation of chemical calculations is key to resolving ion-specificity in these problems. The
coupling of structure with thermodynamics  helps to understand the
molecular mechanisms.  

In the following, we first discuss basic aspects of QCT, including
surface potentials relevant to ion hydration calculations involving
interfaces, and application of QCT to protein binding sites. Then, we survey our
broad set of simulation results on hydration structures for relevant
metal ions in water. Next, we present our
results on ion hydration free energies, obtained by analyzing local
hydration structures in the QCT formulation. Finally,
we compare local ion hydration structures in liquid water with 
local ion solvation structures in protein binding sites as a structural test of the hydration mimicry concept.  We reserve for
future work the 
 comparison of ion solvation free energies in channel binding sites 
to ion hydration free energies.
That comparison will take into account the environment surrounding the binding 
sites, including the possibility that
the environment constrains local binding site structure \cite{Varma2007,
Varma:2011ho,Rogers:2011}.


\section{QUASI-CHEMICAL THEORY}\label{sec:QCT}
QCT was 
developed \cite{redbook,%
Asthagiri:2010,Rogers:2011}
for just the problems considered here: interaction free energies of
specific ions in 
solutions and protein binding sites \cite{Stevens:2016,Asthagiri:2004hq,Chaudhari:2014wb,Asthagiri:2010,Rogers:2011,Asthagiri:2003je,
Jiao:2012,Dudev2013}. 
The excess chemical potential 
\begin{eqnarray}
\mu_{\mathrm{M}^{q+}}^{(\mathrm{ex})} = \mu_{\mathrm{M}^{q+}}  - kT \ln \rho_{\mathrm{M}^{q+}} \Lambda_{\mathrm{M}^{q+}}{}^3
\label{eq:ideal}
\end{eqnarray}
is obtained from the full chemical potential
less the ideal contribution indicated. Here $T$
is the temperature, $\rho_{\mathrm{M}^{q+}}$ the number 
density of the ion of interest,  and
$\Lambda_{\mathrm{M}^{q+}}$ is the thermal deBroglie
wavelength \cite{Beck:2006wp} of species $\mathrm{M}^{q+}$(aq).  This
interaction free energy analysis can also provide 
\begin{eqnarray}
\mu_{\mathrm{M}^{q+}}^{(\mathrm{ex})}\left(\mathbf{r}\right) =
\mu_{\mathrm{M}^{q+}}  - kT 
\ln \rho_{\mathrm{M}}\left(\mathbf{r}\right) \Lambda_{\mathrm{M}^{q+}}{}^3~,
\label{eq:idealprime}
\end{eqnarray}
describing binding at locations 
$\mathbf{r}$, generally \cite{Beck:2006wp}.

\subsection{Inner-shell Clusters}The physical concepts 
underlying QCT develop from consideration of 
association equilibria
\begin{equation} 
n \mathrm{H_2O} + \mathrm{M}^{q+} \rightleftharpoons \mathrm{(H_2O)}_n\mathrm{M}^{q+}~.
\label{eq:2charged}
\end{equation}
The populations of the 
clusters $\mathrm{(H_2O)}_n{\mathrm{M}^{q+}}$  are identified by a clustering
algorithm, according to which proximal ligands of a 
specific $\mathrm{M}^{q+}$ are
defined as  `\emph{inner-shell}' partners of that ion.  The
theory develops by treating the cluster         
$\mathrm{(H_2O)}_n{\mathrm{M}^{q+}}$ as a molecular 
component of the system.   

QCT is then a fully elaborated statistical mechanical theory
that activates modern molecular 
computation \cite{redbook,%
Asthagiri:2010,Rogers:2011}.
What is more, QCT can be closely coordinated with molecule simulation
calculations. In that way, QCT provides a
compelling molecular theory of liquid
water itself \cite{shah2007balancing}.

Thus, application of QCT  begins with identification of
inner-shell configurations of an ion of interest. A simple procedure is
to identify those water molecules with O atoms within a distance
$\lambda$ from a metal ion as inner-shell partners.  From there, with $n$
water ligands in the cluster, the free energy is 
elaborated as 
\begin{eqnarray}
\mu^{\mathrm{(ex)}}_{\mathrm{M}^{q+}} = -kT\ln K^{(0)}_{n}\rho_{\mathrm{H_2O}}{}^{n} 
		+kT\ln p_{\mathrm{M}^{q+}}(n)  
	+\left(
	\mu^{\mathrm{(ex)}}_{\mathrm{(H_2O)}_n{\mathrm{M}^{q+}}}-n\mu^{\mathrm{(ex)}}_{\mathrm{H_2O}}
	\right)~,
\label{eq:1}
\end{eqnarray}
without statistical mechanical approximation.  This
formula is correct for any physical choices of $\lambda$ and $n$. 
Figure~\ref{fig:QCT} guides us through the several terms of 
Eq.~\eqref{eq:1}, which we discuss in the following.

\begin{marginnote}[]
\entry{QCT}{QCT partitions the free energy into three 
distinct contributions based on a clustering algorithm.}
\end{marginnote}

The clustering free energy for 
Eq.~\eqref{eq:2charged} is based upon the equilibrium ratio
\begin{eqnarray}
K_n =
\frac{p_{\mathrm{M}^{q+}}(n)}{p_{\mathrm{M}^{q+}}(0)\rho_{\mathrm{H_2O}}{}^n}~.
\label{eq:Kratio}
\end{eqnarray}
The factor $K^{(0)}_{n}$ appearing in Eq.~\eqref{eq:1} is that
equilibrium constant $K_n$ evaluated for the case that the 
external
medium is an ideal gas. Evaluation of $K^{(0)}_{n}$ is 
accessible
with widely available tools of molecular computational 
chemistry and can be validated against high pressure
mass spectrometry data \cite{Peschke1998,%
tissandier1998proton,Kebarle:1977uu,Keesee:1980ti,%
Keesee:1980fj,Castleman:1986fu,CastlemanJr:1988vo}.

The ideal gas characteristic $K^{(0)}_n$  
conventionally takes  $p=1$~atm, thus identifying 
the ideal density $p/RT$ with 
$p=1$~atm.  Our applications target $\rho_{\mathrm{H_2O}}= 1$ gm/cm$^3$
as the density of liquid water at $T=298$~K and $p=1$~atm. Then
$\rho_{\mathrm{H_2O}}RT \approx  1354$~atm. These 
density factors describe the availability of water for binding the ion. 
In the application to ion hydration, this availability is enhanced
by 1354 relative to the ideal 
$p=1$~atm value.

Practical calculations of the outer-shell 
free energy term $\left(
	\mu^{\mathrm{(ex)}}_{\mathrm{(H_2O)}_n{\mathrm{M}^{q+}}}-n\mu^{\mathrm{(ex)}}_{\mathrm{H_2O}}
	\right)$ are set by
adopting a statistical thermodynamic model of the environment
of the $\mathrm{(H_2O)}_n{\mathrm{M}^{q+}}$
cluster for the incipient free energy 
balance of Eq.~\eqref{eq:1}
Here, we employ the  polarizable continuum model
(PCM) \cite{Tomasi:2005tc}. With PCM, the external boundary of 
the cluster-solute is
defined by spheres centered on each of the atoms. 
PCM results are
sensitive to the values of the  radii. But   sensitivity to the values of the radii often
balances-out in the free energy difference when the ion is 
buried by the ligands.  

Finally, Figure~\ref{fig:QCT} identifies the contribution 
$kT\ln p_{\mathrm{M}}(n)$ 
with release of the 
constraint requiring $n$ waters in ion association. 
The left-side of Eq.~\eqref{eq:1} 
is independent
of $n$, so the complement  provides
$kT\ln p_{\mathrm{M}}(n)$ to within a constant.  Considering 
a specific $\lambda$, the minimum value of $kT\ln p_{\mathrm{M}}(n)$ 
identifies the
most probable $n$, which we denote by $\bar n$.   Then we drop that
statistical contribution
\begin{eqnarray}
\mu^{\mathrm{(ex)}}_{\mathrm{M}^{q+}} \approx -kT\ln K^{(0)}_{\bar n}\rho_{\mathrm{H_2O}}{}^{\bar n}
	+\left(
	\mu^{\mathrm{(ex)}}_{\mathrm{(H_2O)}_{\bar n}{\mathrm{M}^{q+}}}-{\bar n}\mu^{\mathrm{(ex)}}_{\mathrm{H_2O}}
	\right)~.
	\label{eq:4}
\end{eqnarray}
Although $\bar n$ minimizes that approximation error, the 
magnitude of the correction can be estimated from simulation results. 

The advantage of QCT is the separation of solvation free energies 
into components from inner-shell and outer-shell solvent molecules, 
addressing the chemical physics issues in analysis 
of clusters.  
Those issues include proper overlap
repulsions, polarizability, charge transfer, London dispersion
interactions, $n$-body ligand interactions generally, distortion of
flexible ligands, and zero-point motion of the cluster.


\subsection{Potential of the Phase \label{SP}}
\begin{figure*}
  \begin{minipage}{0.50\textwidth}
	\includegraphics[width=0.95\textwidth]{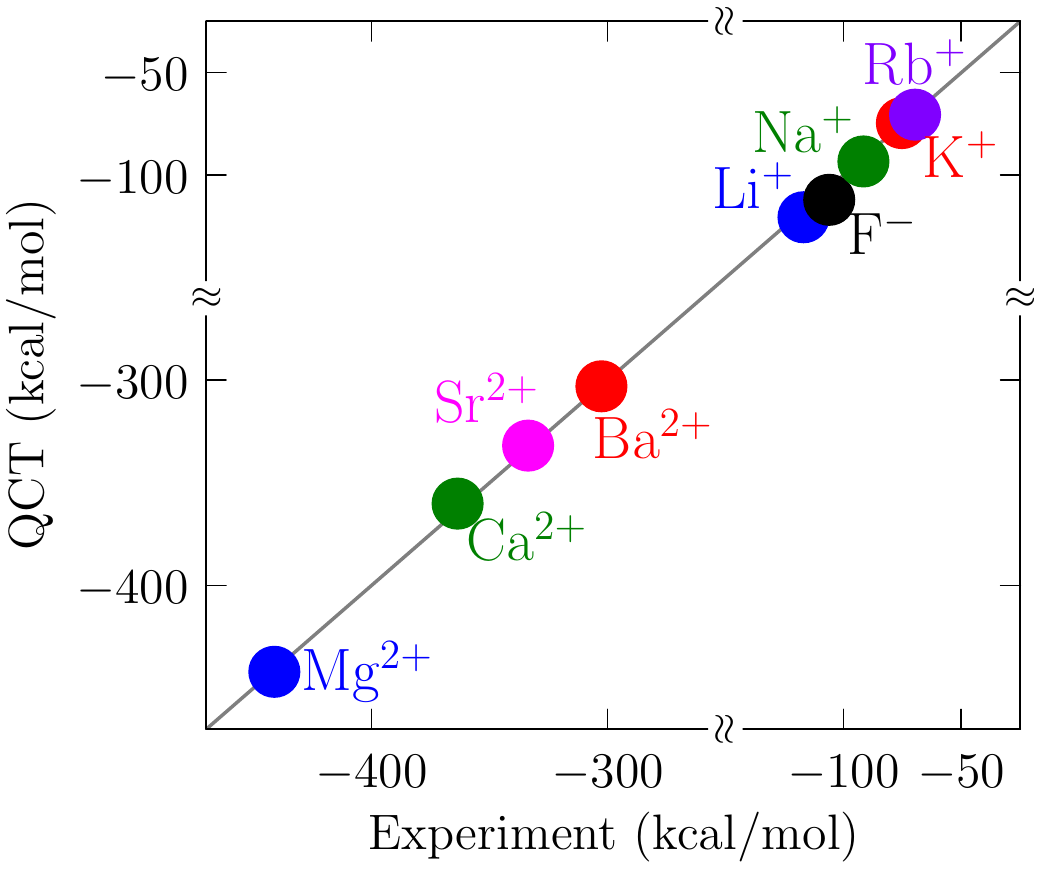}
	\hfill
	\end{minipage}
	  \begin{minipage}{0.48\textwidth}
	\includegraphics[width=0.95\textwidth]{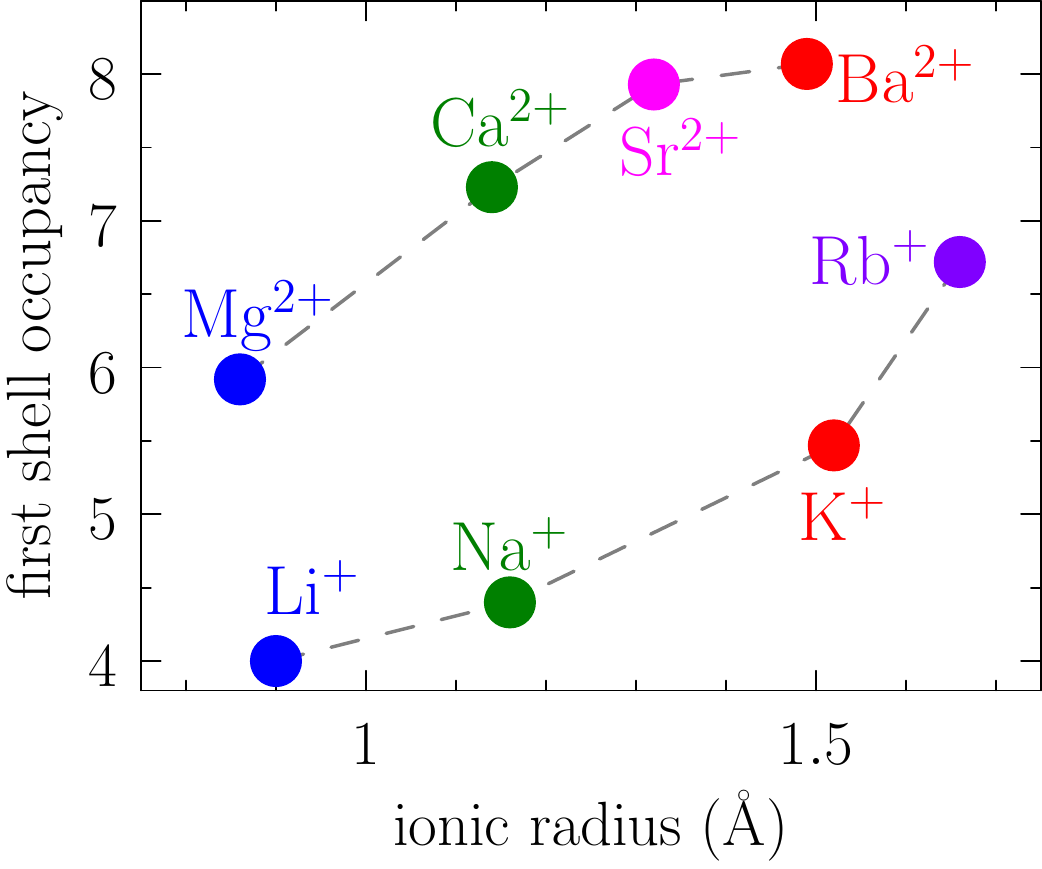}
	\end{minipage}
	\caption{Left Panel: Comparison between hydration free 
	energies calculated
	from QCT and the experimental tabulation of
	Ref.\citenum{Marcus:1994ci}, except for the F$^-$ value 
	which was updated in Ref.~\citenum{marcus2015ions}.
	Note that internal consistency of free energies
	for each collection of identically charged ions supports the view that any surface potential is treated reasonably. Right Panel: Occupancy of the nearest hydration shell defined by
	the minima in RDF
	(Figure~\ref{fig:gr}) as a
	function of ionic radius. Water occupancy 
	increases with increasing ion size, and water occupancy
	increases for divalent ions compared to monovalent ions of
	similar size.}
	\label{fig:exp_QCT}
\end{figure*}

Discussion of surface potentials brings forward the 
subtlety of \emph{experimental} testing 
of computed single-ion free energies (Figure~\ref{fig:exp_QCT}).
To that end, we augment the free energies of Eq.~\eqref{eq:idealprime}
according to
\begin{eqnarray}
\mu_{\mathrm{M}^{q+}}  = q e \Phi 
+  kT \ln
\rho_{\mathrm{M}^{q+}} \Lambda_{\mathrm{M}^{q+}}^3 + 
\mu_{\mathrm{M}^{q+}} ^{(\mathrm{ex})} ~,
\label{eq:electrochempot}
\end{eqnarray}
by the electrostatic contribution, $q e \Phi$, for each of the
conducting phases considered \cite{pethica2007electrostatic}.
Since $\Phi$ does not depend on chemical details of the
ion, this extension plays no role in assessing
the free energy of neutral combinations of ions.  $\Phi$ does not 
depend on the ion
size or structure, nor the distribution of electric charge within the ion. 
We will call $\Phi$ the ``\emph{potential 
of the phase}.''  That quantity is determined through  Poisson's equation of
electrostatics, with charge densities for the generally heterogeneous
system and boundary conditions. $\Phi$  thus 
depends on  conditions bounding and external to the 
phase.  Considering a \emph{conducting} 
homogeneous fluid phase, $\Phi$ is a constant
throughout \cite{LLV8}. The bulk composition is
charge-neutral for such a phase.  

For the case of two conducting fluid phases in equilibrium with respect
to ion transfer between the phases, the difference $\Delta \Phi$ --- the
contact or junction potential \cite{Zhou:1988fz} --- can be tied to
conditions of transfer equilibrium of ions and the neutrality of the
bulk 
compositions \cite{Zhou:1988fz,Nichols:1984ev,Pratt:1992tg,%
you2014comparison}.
For a $q$-$q$ electrolyte MX,
\begin{eqnarray}
		2  q e \Delta \Phi = -\Delta 
\left\lbrack\mu_{\mathrm{M}^{q+}}^{(\mathrm{ex})}-\mu_{\mathrm{X}^{q^-}}^{(\mathrm{ex})}\right\rbrack ~.
\label{eq:surfpot}
\end{eqnarray}
The right side characteristics need not address the
interface between the two phases, but are aspects of the bulk solutions 
for the case $\Phi = 0$ --- or some other fiducial value --- for each 
phase  considered individually, as is clearly permissible.  

Though the spatial 
transition of $\Phi$ through an interfacial region
from one homogeneous conducting solution to the other is less simple,  the simple result Eq.~\eqref{eq:surfpot} was
emphasized many years
ago \cite{Zhou:1988fz,Nichols:1984ev,Pratt:1992tg,you2014comparison}
``\ldots  the junction potential in the ideal solution limits differ in
general from the pure-solvent values'' \cite{Zhou:1988fz}. Further,
``since all real polar solvents are to some extent ionized, one has to
take this effect into account in relating theory to
experiments'' \cite{Zhou:1988fz}. This situation does not seem to be
widely appreciated.

There is  broad interest in the idealized case of the fluid electrolyte
solution bounded by a dielectric (non-conductor), perhaps a vacuum. 
This is the context for discussions of a ``\emph{surface
potential}'' \cite{Lyklema:2017vy}. The solution, being a conductor, 
will still exhibit a
spatially constant $\Phi$.  But the electrostatic potential will vary
through regions bounding and exterior to the solution.  The change in the
electrostatic potential with passage out of the solution then
requires further specification. Further 
theoretical modeling specifications are simpler if a 
sub-macroscopic cavity is  
imposed \emph{internally}.  Then the
contacts of the solution with the cavity are
studied \cite{Beck:2013gp,Pollard:2018hi}. Questions remaining
include whether changes in the electrostatic potential are
satisfactorily independent of cavity size and whether 
the electrostatic potential might be spatially constant 
enough to serve in the thermodynamic 
formulation Eq.~\eqref{eq:electrochempot} for ions with 
different distributions of charge.  

\begin{marginnote}[]
\entry{Surface potential}{Predictions of 
differences between ions of the same charge are 
unaffected by a surface potential contribution.}
\end{marginnote}

Reserving such issues for 
future research \cite{doyle2019}, 
we finally discuss the relevance of the
QCT approach, based upon Eq.~\eqref{eq:4} Since 
Eq.~\eqref{eq:2charged} is balanced with respect to
\emph{charge}, the $K_n$ of Eq.~\eqref{eq:Kratio} do not
involve the potential of the phase \cite{Beck:2006wp}.
The examination of the 
TATB hypothesis \cite{Lesniewski:2018jd,Pollard:2018hi,%
Marcus:1987hi,Schurhammer:2000hv,%
Duignan:2018hl,Leung:2009dx},
and Marcus's modeling of his tabulated values \cite{Marcus:1987hi,Marcus:1994ci} 
to depend quadratically on ion charge,  make those values 
natural for comparison with the QCT single-ion free 
energies  
that use the PCM here for the cluster free energy of 
Eq.~\eqref{eq:4} That the values of an 
alternative tabulation \cite{friedman1973thermodynamics} are 
distinctly different is a cautionary point.

    \subsection{QCT and Protein Binding Sites}

 
Of special interest are the ion free energies 
$\mu_{\mathrm{M}^{q+}}^{(\mathrm{ex})}\left(\mathbf{r}\right)$
at a defined binding site, as anticipated with 
Eq.~\eqref{eq:idealprime} and Figure~\ref{fig:Hydmim}.  
QCT, being a general approach, is applicable to those 
problems \cite{Varma2007,Stevens:2016,chaudhari2018SrBa,rossi} though QCT approaches 
deserve further development and detailed refinement 
for that context.

Here we use the  
context of \Ba\ occupying the innermost binding 
site of K-channel 
(S4)  to  
exemplify one approach.  
In the example, we start with the K-channel 
crystal structure (PDB 1K4C) and the S4 site occupied by \Ba.  
Four threonine (THR) amino acids interact with \Ba\ in 
bidentate fashion. Therefore, 
we build Ba(Thr)$_{4}{}^{2+}$ clusters, and, in fact
an energy-optimized cluster structure shows all Thr O atoms
displaced less than 1~\AA\ from the crystal structure for 
the occupied binding site.    This result is consistent with 
a physical intuition underlying QCT;   
namely \cite{Sabo:2013gs,Asthagiri:2010,Chaudhari:2017gs}, 
that interactions of 
metal ions with near-neighbors are localized, hugely 
stabilizing, and matters of first importance.

QCT addresses the solvation free energy of \Ba\ on the basis 
of analysis of clusters 
\begin{eqnarray}
\mathrm{Ba}\left(\mathrm{H}_2\mathrm{O} \right)_m
\left(\mathrm{Thr} \right)_{n-1}{}^{2+} + \mathrm{Thr} 
\rightleftharpoons 
\mathrm{Ba}\left(\mathrm{H}_2\mathrm{O} \right)_{m-2}
\left(\mathrm{Thr} \right)_{n}{}^{2+} + 2\mathrm{H}_2\mathrm{O}~.
\label{eq:bsite}
\end{eqnarray}
that may form. Note that Eq.~\eqref{eq:bsite} makes the 
reasonable assumption that two $\mathrm{H}_2\mathrm{O}$
molecules most naturally replace one Thr ligand.  The equilibria
Eq.~\eqref{eq:bsite},
$n = 1, \ldots , 4$, describe the formation of a binding 
site encapsulating
a \Ba\ ion.
\begin{marginnote}[]
\entry{Stability of protein binding sites to 
changes of water activity}{What are the 
probable $\mathrm{Ba}\left(\mathrm{H}_2\mathrm{O} \right)_m
\left(\mathrm{Thr} \right)_{n-1}{}^{2+}$  coordination cases 
for given Thr solution concentrations? }
\end{marginnote}
From the beginning ($n=1$) state to a final ($n=4$) 
state, this process converts hydrated \Ba\ ions to \Ba\ ions centering a
model binding site.  The free energy change for this process 
depends on the concentrations of the species involved, \emph{i.e.},
a full description requires considering a standard state.  We noted
above with Eqs.~\eqref{eq:1} and \eqref{eq:Kratio} that QCT 
properly resolves a standard state.

The equilibria Eq.~\eqref{eq:bsite} can organize a
solution chemistry experiment involving an aqueous 
solution of a \Ba\ electrolyte as a medium for dissolution 
of Thr.  The dissolved concentration of Thr would 
be tracked, and the coordination of \Ba\ by the 
Thr would be interrogated.   The question `\emph{what are the 
probable $\mathrm{Ba}\left(\mathrm{H}_2\mathrm{O} \right)_m
\left(\mathrm{Thr} \right)_{n-1}{}^{2+}$  coordination cases 
for given Thr solution concentrations}?' would naturally arise.
Alternatively expressed: from the Thr completion side of the 
scheme Eq.~\eqref{eq:bsite}, how much would the 
H$_2$O activity have to increase, by dilution of the Thr,
for H$_2$O molecules to disrupt the (Thr)$_4$ binding site?  
Though this solution chemistry experiment would be an exceedingly natural 
assessment of biomolecular hydration mimicry,  
it has not been done as far as we know.  Answers to such questions, and 
the natural follow-up questions, would 
address current  issues of H$_2$O occupancy 
of potassium 
channels \cite{tilegenova2019structure,Oster:2019wq,%
Hummer:2014jj,%
kopec2019molecular}.

\section{ION HYDRATION STRUCTURE}\label{sec:hydra_struc}

\begin{figure*}
\centering
\begin{minipage}{0.49\textwidth}
\includegraphics[height=0.21\textheight]{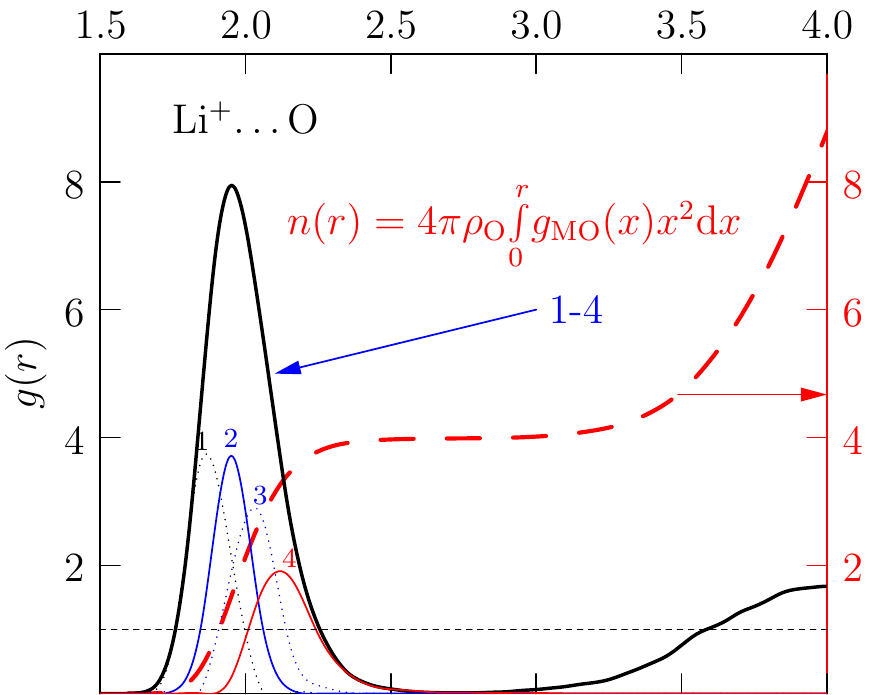}
\end{minipage} \hfill
\begin{minipage}{0.49\textwidth}
\includegraphics[height=0.21\textheight]{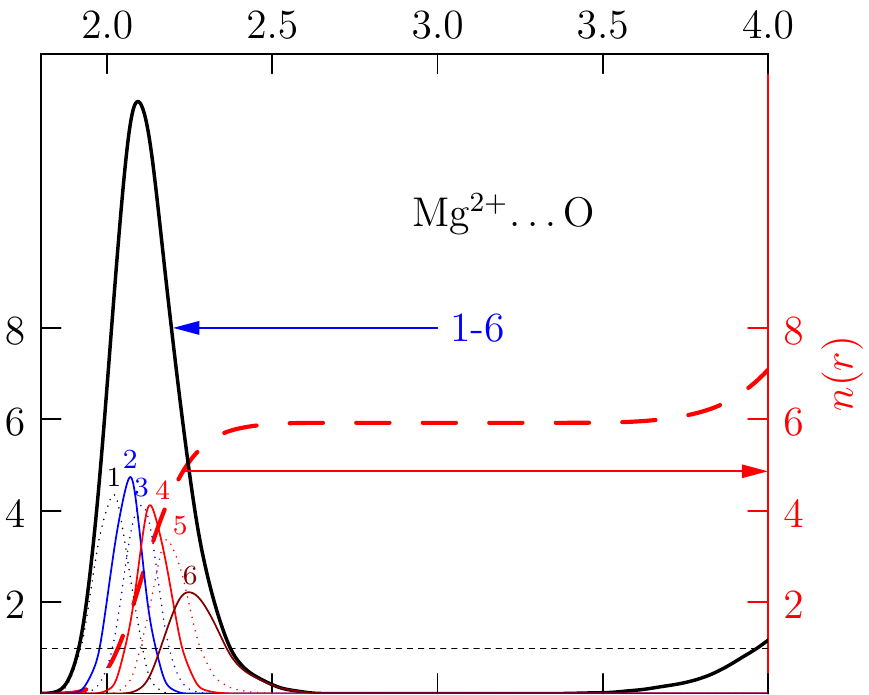}
\end{minipage}
\begin{minipage}{0.49\textwidth}
\includegraphics[height=0.21\textheight]{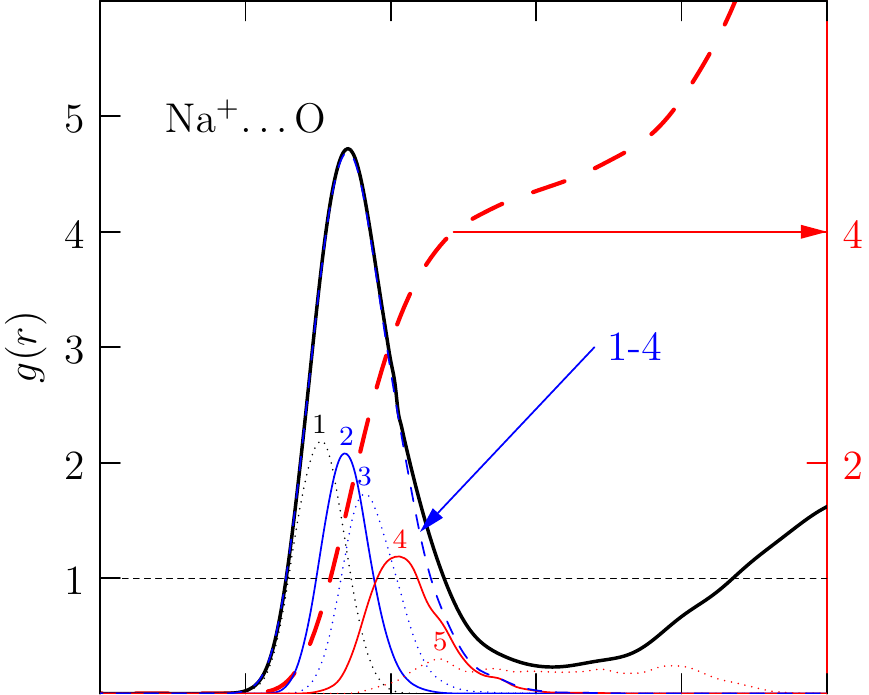}
\end{minipage}\hfill
\begin{minipage}{0.49\textwidth}
\includegraphics[height=0.21\textheight]{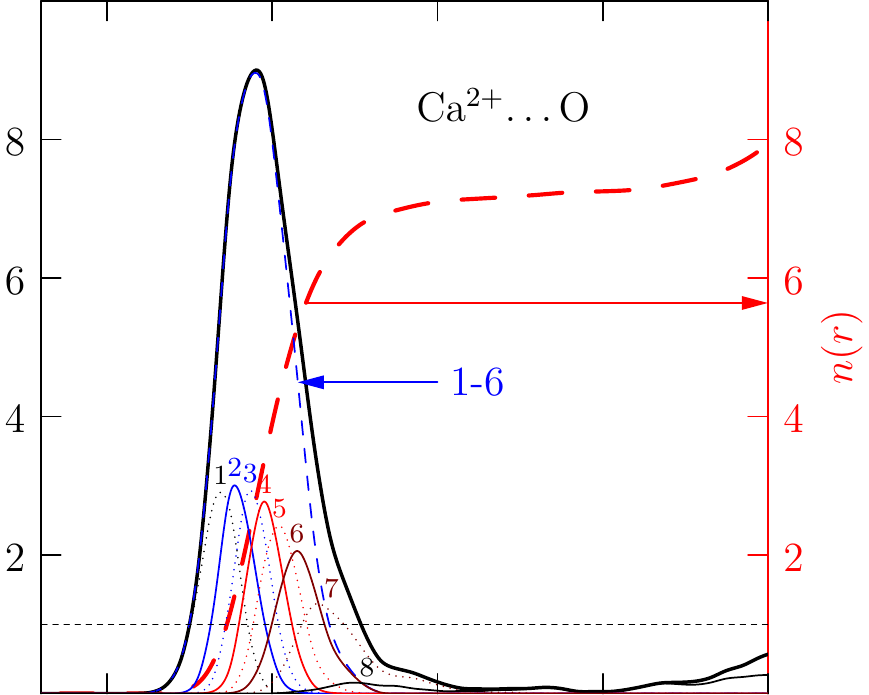} 
\end{minipage}
\begin{minipage}{0.49\textwidth}
\includegraphics[height=0.21\textheight]{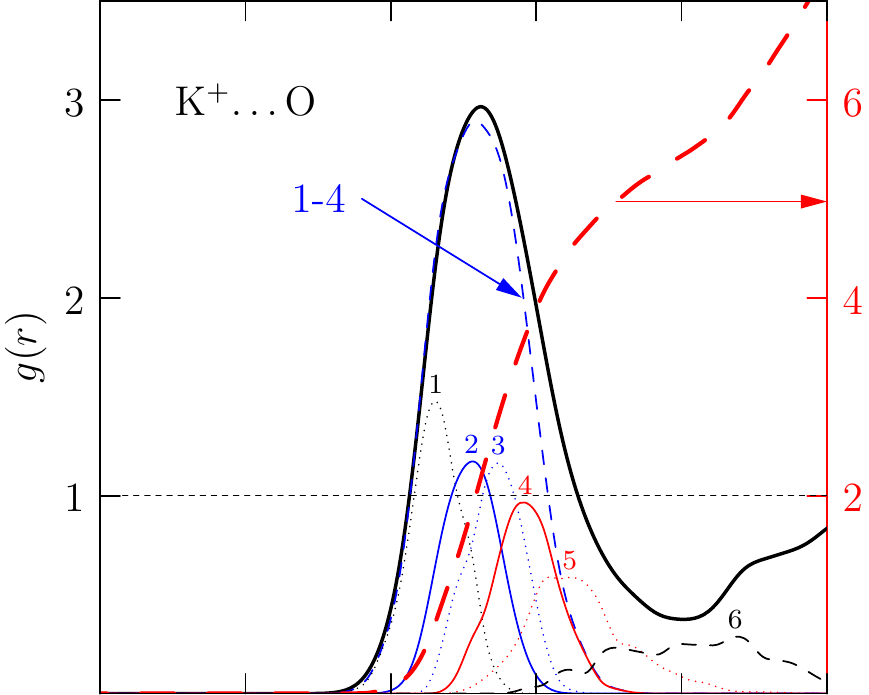}
\end{minipage}\hfill
\begin{minipage}{0.49\textwidth}
\includegraphics[height=0.21\textheight]{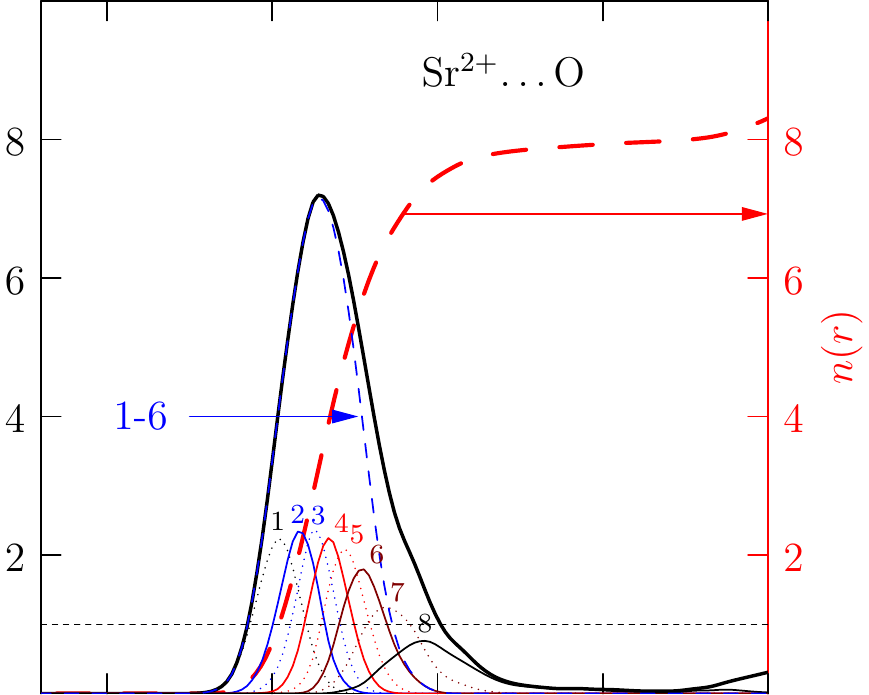} 
\end{minipage}
\begin{minipage}{0.49\textwidth}
\includegraphics[height=0.21\textheight]{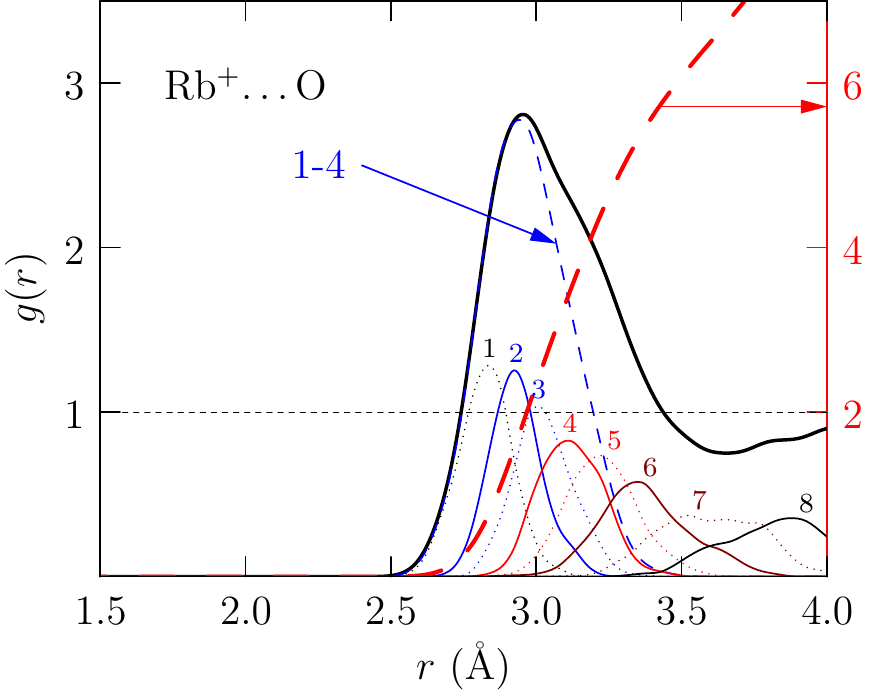}
\end{minipage}\hfill
\begin{minipage}{0.49\textwidth}
\includegraphics[height=0.21\textheight]{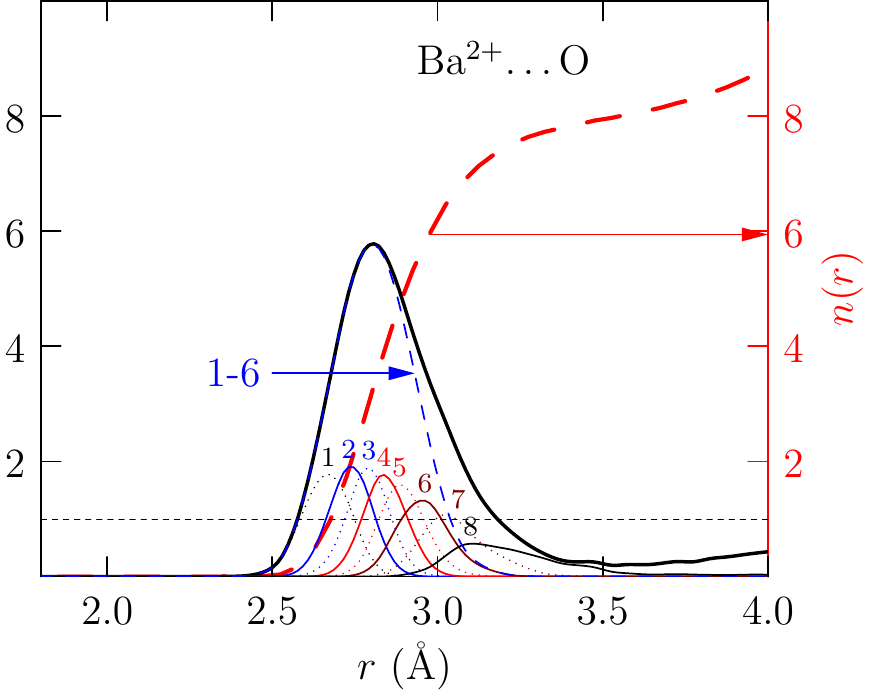}
\end{minipage}
\caption{AIMD radial distributions of water O around eight metal ions, together
with neighborship-ordered
distributions \cite{Mazur:1992gg,zhu2011generalizations}. Red dashed
curves are coordination numbers. 
Peak positions are given in Table~\ref{table:0}.}
\label{fig:gr}	
\end{figure*}

Characterization of local ion hydration structure 
provides data essential for evaluation of
the hydration mimicry concept. 
Radial distributions (RDFs: Figure~\ref{fig:gr}) of O (water)
obtained from AIMD simulations for eight metal ions in water 
define inner-shell structures 
for alkali metal ions and alkaline earth metal ions. 
Those AIMD results are consistent with the accurate experimental 
determinations of peak 
positions (Table~\ref{table:0}), and 
with alternative studies of particular 
cases \cite{Jiao:2006ik,%
Mehandzhiyski:2015gu,Soniat:2015br}.  
These ions are often 
compared to investigate effects of the
ion charges and sizes on channel behavior. For the 
smaller ions, there 
is an obvious separation between inner and outer hydration
shells.  As the ion-size increases from
Li$^+$ to Rb$^+$ and \Mg to \Ba  (Figure~\ref{fig:gr}), the 
mean inner-shell occupancies  become less distinct.

Neighborship decompositions \cite{Mazur:1992gg,zhu2011generalizations}
of those distributions characterize the natural hydration 
shell structure \cite{Varma2007,Chaudhari:2017gs}. For
the monovalent ions considered, the nearest four  water
molecules suffice to describe the RDF maxima even though the total
occupancy of the inner shell increases with the ion size
(Figure~\ref{fig:exp_QCT}). Six   water molecules suffice for the RDF
maxima for the alkaline earth ions, but the inner-shell occupancies
reach six or eight (8). 

Note that sometimes a neighbor distribution can be multi-modal.
See, for example, the case of $n =6$ for K$^+$(aq) (Figure~\ref{fig:gr}).
We call these `\emph{split-shell}' cases.   Since the 
basic development, Eq.~\eqref{eq:1}, is correct independently of  $\lambda$ and $n$, we could agree 
to limit QCT applications to non-split-shell cases of 
$\lambda$ and $n$, or we could implement the more 
involved computational work to treat the multi-modal occupancies 
directly \cite{ajay_cl}.

\begin{marginnote}[]
\entry{Split-shell coordination cases}{The 
6-th nearest neighbor of K$^+$(aq) 
sometimes occupies an inner-most shell
and, at other times, outer hydration shells.}
\end{marginnote}


While hydration properties of the monovalent ions have been studied extensively, fewer works have reported on the hydration properties of divalent ions.
\Mg\ hydration has long been of biophysical interest. Dudev
\emph{et al.}\cite{Dudev:1999cn} used electronic structure calculations
to evaluate energy changes for replacing water molecules with other
ligands. Six water molecules established the nominal coordination number
for Mg$^{2+}{(\mathrm{aq})}$ ion \cite{Dudev2013}. 
Such results are consistent with the X-ray
diffraction measurements on MgCl$_2$ solution \cite{Caminiti:1979ci} and
our AIMD results (Figure~\ref{fig:gr}).

Our RDF for \Ca~ion, and its neighborship decomposition (Figure~\ref{fig:gr}),
suggests that the eighth
water molecule splits occupancy between inner and outer hydration
shells, giving a hydration number between seven and eight (7-8).   In comparison, Marcus \cite{Marcus:1994ci} and 
Dudev \emph{et al.} \cite{Dudev2013},
predicted a hydration number of 7 for \Ca(aq). That possibility 
deserves further investigation. 

Previous AIMD simulations \cite{Harris:2003jq,Spohr:2002kn} calculated
a mean inner-shell occupancy of 7.5 for \Sr(aq),  lower than values of 9.3
and 14.9 \cite{Ohtaki1993} from experimental work.  Our result (7.9)
supports the previous simulation effort.  Although simulation and
experiment differ in this respect, our  RDF peak positions closely match 
the experimental results, and predictions of hydration free
energy match experimental values (see below), suggesting a reliable 
prediction of hydration number.  

An octa-coordinated hydration structure for \Ba(aq) was proposed in
1933 \cite{bernal}.  Though a several studies 
have interceded since, our result here lends support to that proposal and 
experimental data.  The  hydration free energy computed 
on that basis matches experimental results, as presented below.

\begin{table*}
\begin{center}
\begin{tabular}{| c | l | l || c | l | l  | }
\hline    
 {Ion$^+$} & AIMD    &  EXP  & {Ion$^{2+}$} &  AIMD    &  EXP  \\
\hline
Li & 1.95\,\cite{Leung:2009dx,alam:li}   & 1.96\,\cite{Mason:2015}  
     & Mg & 2.09\,\cite{Chaudhari:2017gs} & 2.04\,\cite{Caminiti:1977ff} \\   
Na & 2.37\,\cite{Varma2006}  & 2.38\,\cite{Na_Fulton} 
     & Ca & 2.45\,\cite{Chaudhari:2017gs} & 2.43\,\cite{Caminiti:1977ff,Licheri:1975gk} \\
K & 2.73\,\cite{Varma2006,Varma2007} & 2.73\,\cite{K_Fulton} 
     & Sr & 2.64\,\cite{chaudhari2018SrBa} & 2.63\,\cite{Ramos:2003wd,Pfund:2002bf} \\
Rb & 2.95\,\cite{Sabo:2013gs} & 3.10\,\cite{Rb_Ramos} 
     & Ba & 2.81\,\cite{Chaudhari:2014wb} & 2.80\,\cite{dangelo}  \\
\hline
\end{tabular}
\bigskip
\caption{Comparison of radial positions in \AA\ of the RDF-maxima observed in AIMD 
simulation (Figure~\ref{fig:gr}) with experimental values.  References
are in parentheses.}	
\label{table:0}
\end{center}
\end{table*}



\section{ION HYDRATION FREE ENERGY}\label{sec:hydra_fenergy}
Ion hydration free energies set a reference value for ion permeation
through protein channels and transporters. Comparisons of hydration free
energies with experimental results provide baseline tests of molecular
statistical thermodynamic theory. For ions, those comparisons raise the
issue that hydration free energies are obtained by manipulation of
neutral material combinations. Thus, tabulated single ion free energies,
used here, incorporate extra-thermodynamic assumptions, and will
disagree somewhat where different assumptions are
used \cite{Marcus:1994ci,friedman1973thermodynamics}.  We respond to this
issue partially here, in two different ways.  

The first partial response is that computed theoretical free energies
can be tested with results for neutral combinations of ions.  That
requires application of the theory to an example
anion \cite{ajay_F,ajay_cl,chaudhari2017quasi} in addition to the mono-atomic metal ions
of primary interest here. To the extent that the theory performs
satisfactorily for the neutral test case, comparisons among different
cations are supported also. Figure~\ref{fig:exp_QCT} gives such an
initial comparison, supported by a QCT calculation for
F$^-$(aq) \cite{chaudhari2017quasi,ajay_F}. Combining results for LiF
produces $-227.5$~kcal/mol (QCT), in fair agreement with experimental
tabulations, which range between $-229$~kcal/mol 
and $-232$~kcal/mol \cite{friedman1973thermodynamics,%
Marcus:1994ci,tissandier1998proton}. This discussion is only a partial
response for molecular theory because it declines advantages from
dissection of a net result into physically meaningful parts, the
single-ion free energies.  

Our second partial response is that the theory should address the
problem sufficiently thoroughly that accuracy of the individuals steps
of the theory can be checked. Developing this second partial response
here provides the opportunity to investigate the ingredients that are
central to the performance of QCT theory.

\begin{figure*}[htb]
    \begin{minipage}{0.49\textwidth}
	\includegraphics[width=\textwidth]{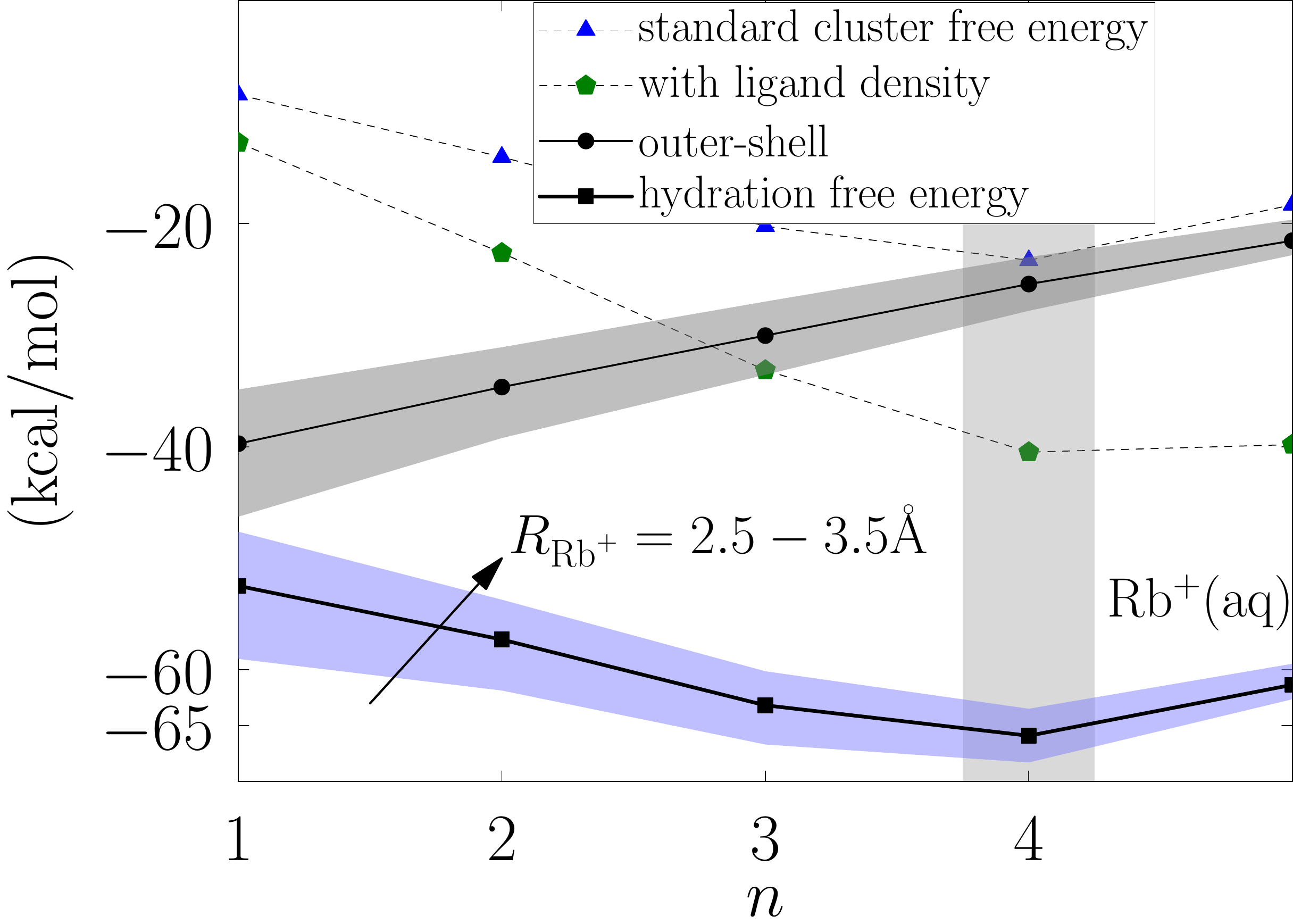}
	\end{minipage} \hfill
    \begin{minipage}{0.49\textwidth}
	\includegraphics[width=\textwidth]{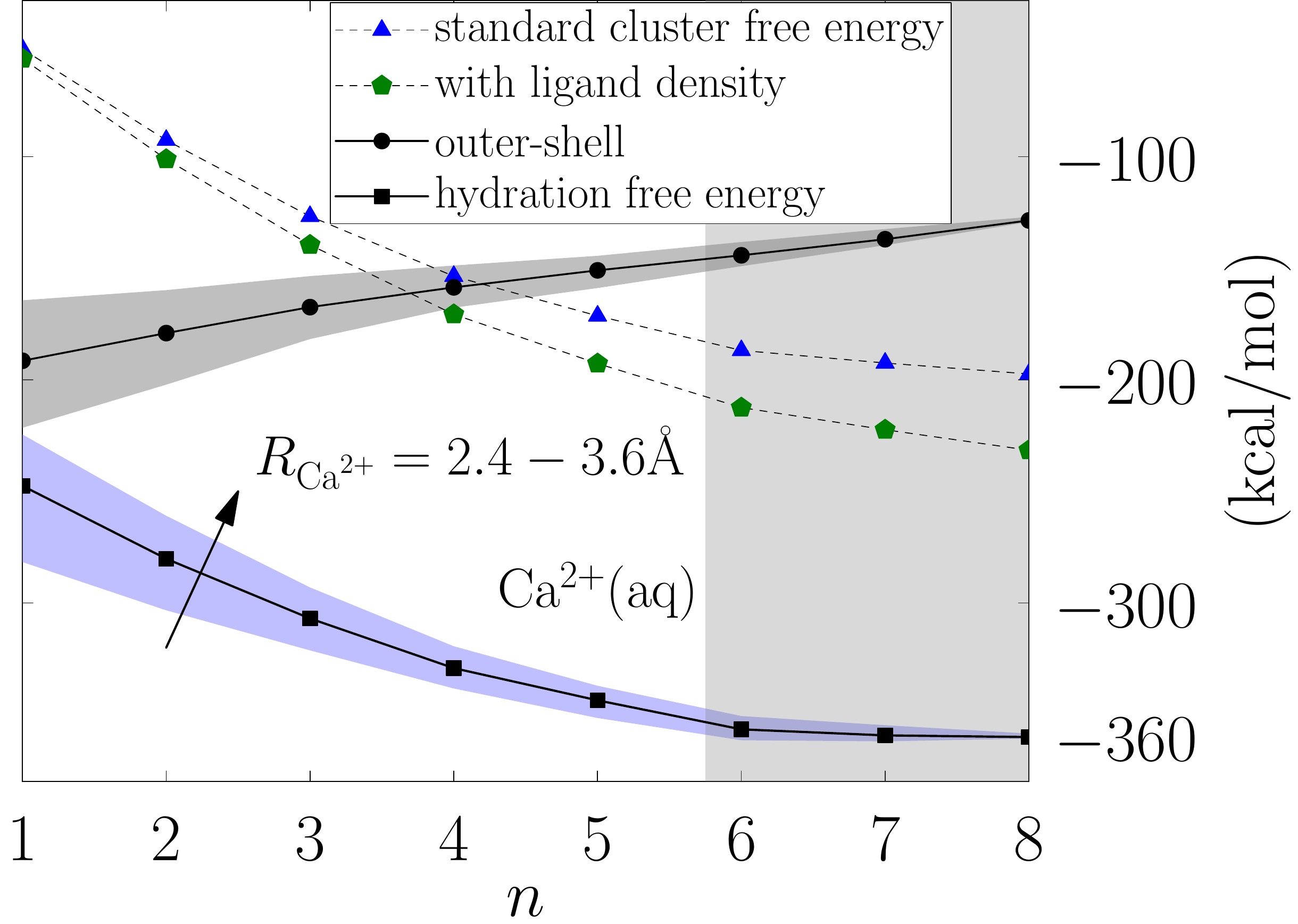}
	\end{minipage}
	\caption{Dependence of individual QCT contributions on
	inner-shell occupancy $n$. Tracking $-kT\ln p(n)$, the net QCT
	combination is minimal for the most probable coordination number
	(Figure~\ref{fig:gr}).  For probable occupancies, the net result is
	insensitive to $R_{\mathrm{M}^{q+}}$ and closely agrees with the
	experimental tabulation of Ref.~\citenum{Marcus:1994ci}. The
	standard cluster free energy contribution at the proper ligand density
	(green pentagons) accounts for more than half the hydration free
	energy.}\label{fig:diff_contri}
\end{figure*}

Considering the several ingredients that are combined to evaluate the
net free energy, we base the present discussion on the examples of
Rb$^+$ and Ca$^{2+}$ (Figure~\ref{fig:diff_contri}).  Those free energy
contributions (Figure~\ref{fig:diff_contri}) are all substantial on a
chemical energy scale; \emph{i.e.}, that is, comparable to traditional
chemical bond energies.

Noting the trends in each free energy contribution
(Figure~\ref{fig:diff_contri}), the gas phase association term becomes
more favorable with increased $n$ for Rb$^+$ and Ca$^{2+}$. The contribution to
\begin{marginnote}[]
\entry{Ion hydration free energies}{
Ion hydration free energies are as large as
chemical bond energies.}
\end{marginnote}
hydration free energy from waters beyond the inner-shell clusters
becomes less favorable as cluster sizes increase.  That outer-shell
contribution also varies with boundaries, explored here for $\lambda$
set between the first maximum and minimum of the RDF for each ion.  The variation
with boundary becomes small upon reaching clusters with full occupancy
of the inner shell, at $n$=4 for Rb$^+$ and $n$=6-8 for Ca$^{2+}$ (vertical gray shading in
Figure~\ref{fig:diff_contri}). 

The variation of the free energy results of Figure~\ref{fig:diff_contri}
with $\lambda$ and $n$ is encouragingly simple.  Nevertheless, the
dielectric model for the outer-shell contributions, here PCM, can be
problematic when $n$ differs substantially from full
occupancy \cite{Sabo:2013gs}. In contrast, the free energies can be
satisfactory when this inner-shell occupancy is saturated because of the
balance between the cluster and ligand terms of the rightmost
contribution of Eq.~\eqref{eq:4}.  Sensitivity to adjustment of
boundaries for dielectric models is moderated then because the adjusted
boundaries are somewhat buried in the cluster, though otherwise
balanced.   This point is significant since continuum models are
typically sensitive to the boundaries, and those
boundaries are not independently defined by physical principle.
Indeed, boundaries \cite{Linder:2004eu} are ``\ldots a complicated function of
density, temperature, and molecular parameters \ldots " Nevertheless, 
the  rightmost (cluster) contribution of Eq.~\eqref{eq:4} is the
principal theoretical approximation in QCT calculations.  

\begin{marginnote}[]
\entry{Insensitivity to cavity radius of the ion}{When 
the inner-shell is full, the cavity
radius set for the ion core is less important.}
\end{marginnote}


\section{LOCAL STRUCTURE COMPARISON}\label{sec:hyd_mim}
We can compare hydration structures  with crystallographic data for
ions in the binding sites of proteins.   
That structure comparison takes us another step in
testing the hydration mimicry concept.  

\begin{figure*}[ht]
\centering
\includegraphics[height=0.88\textheight]{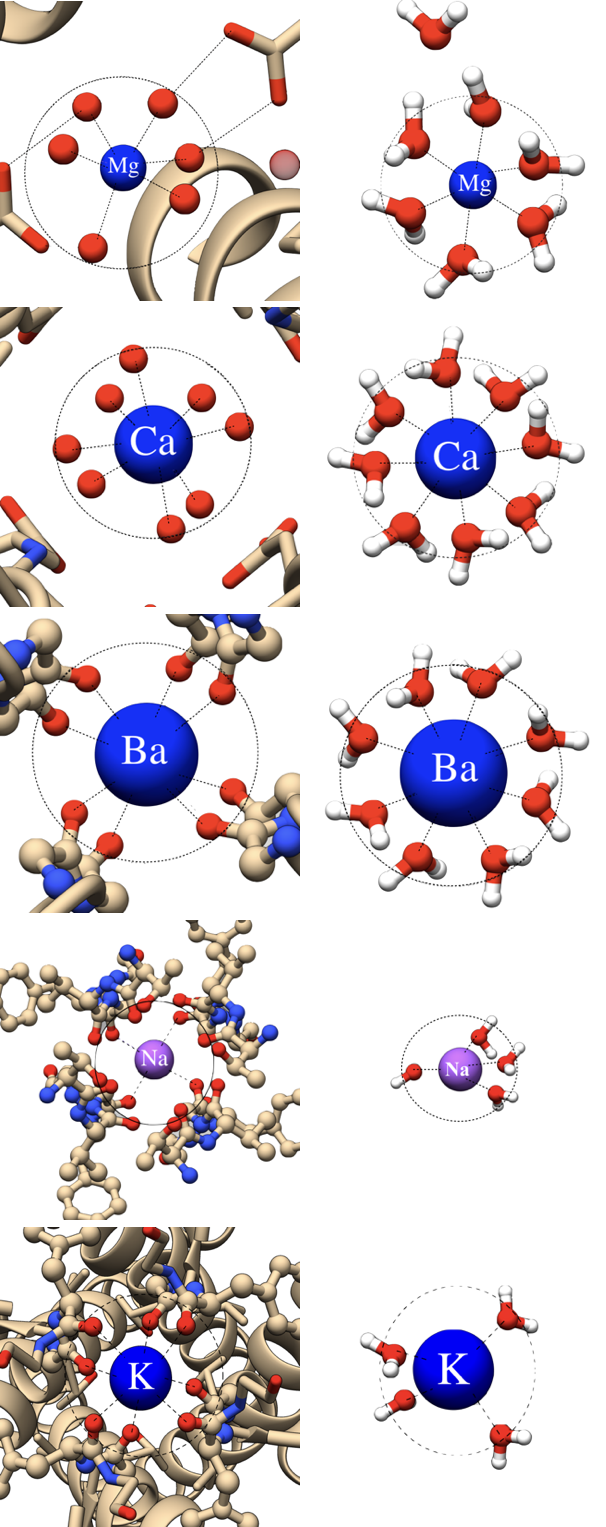}
%
%
%
%
        \caption{Comparison of solution optimized ion-water clusters 
        (right) with 
        X-ray crystal structures of ions
		inside the channels (PDB, left top to bottom,
		4U9L\cite{Payandeh2013},  
		4PDR\cite{Jiang:2014}, and 1K4C\cite{Zhou:2001vo} for the bottom three). Except for the K$^+$ case, the 
		ion coordination inside the 
		channel mimics the near-neighbor hydration structure, 
		consistent with  the  
		hydration mimicry idea.}\label{fig:Hydmim_ion}
	\end{figure*}

We begin with proteins permeable to \Mg\ and
\Ca. Crystallographic data are available for recently discovered
Mg-transporter structures \cite{Payandeh2013,Takeda2014,Dudev2013}.
Compare the \Mg~ion in the Mg-transporter binding site\cite{Takeda2014}
(Figure \ref{fig:Hydmim_ion})  with a PCM-optimized structure of the
$\left\lbrack \mathrm{Mg}\left(\mathrm{H}_2\mathrm{O}\right)\right\rbrack_{6+1}{}^{2+}$,
composed of six waters occupying the inner solvation shell and one water
in the second shell. In bulk solution (Figure~\ref{fig:gr}),
six inner-shell waters establish the near-neighbor distance of 2.1~\AA. 
Unlike K-channels, where K$^+$ ions directly contact protein oxygen
atoms, \Mg\ carries along its aqueous inner-shell as it occupies the
protein binding site. 

Similarly, the Ca$_\mathrm{v}\mathrm{Ab}$ structure shows a fully
hydrated Ca$^{2+}$ ion coordinated by 8 water molecules \cite{Tang2014}.
These water molecules further coordinate with aspartate and asparagine
side chain oxygens of the channel in the outer solvation environment. In
aqueous solution, six waters establish the near-neighbor distance at
2.5~\AA, though up to eight waters can be considered in broader
settings (Figure~\ref{fig:diff_contri}).   

Thus (Figure~\ref{fig:Hydmim_ion}), the divalent ions observed in the
crystal structures of \Ca and \Mg selective binding sites bind directly
with local water molecules \cite{Wilson:2011he}. This
hydration mimickry is nearly perfect and longer-ranged effects come
to the foreground with further selectivity.

In the K-selective KcsA channel, K$^+$ ions can occupy 
each of the  four binding sites
(S1-S4) in the selectivity filter, coordinated by eight oxygens from carbonyl groups either from the protein backbone or from threonine side
chains (Figure~\ref{fig:Hydmim_ion}) \cite{MacKinnon:2003ca}.  The
average distance between oxygen atoms and K$^+$, the cavity size, is
2.8~\AA.  No water molecules contact K$^+$ in this case.  

In comparison, AIMD studies of K$^+$ hydration structure show occupancy
in the inner shell by 4-6 waters (Figure~\ref{fig:gr}). The peak in the
RDF saturates with four  waters, located at a distance of 2.7~\AA\ from
the ion, though six waters can be considered in broader settings 
(Figure~\ref{fig:diff_contri}). The sixth water splits occupancy between inner
and outer shells. In contrast to the eight (8) ligating oxygens in the
KcsA crystal structure, in bulk solution the eighth water of hydration
occupies the outer solvation volume. 

The  differences in K$^+$contacts  raise several questions. Firstly, 
why is the number of K$^+$
inner-shell O atoms less in water than in the KcsA binding site?
Secondly, how can moderate transfer free energy profiles
(Figure~\ref{fig:Hydmim}) arise when local structures differ?  Thirdly,
can local contacts account for selective binding of K$^+$ over Na$^+$?
Prior works have addressed these questions
\cite{Varma2007,Varma:2011ho}, and we expect they will be reviewed 
further in upcoming works.

Since K-selective KcsA channels select against Na$^+$ ions by a factor
of 1000:1, the rejected ions seldom competitively occupy the KcsA
pore. But in the absence of K$^+$, Na$^+$ binds to a site
in-between the K$^+$ binding sites \cite{asthagiri2006role}.
The Na$^+$ ions bind in-between sites S3 and S4,
coordinated by \emph{four} carbonyl oxygens arranged in a 
planar geometry and
separated from the ion by 2.4\AA.

AIMD studies indeed show Na$^+$ hydrated by 4-5 
waters (Figure~\ref{fig:gr}). The fifth water splits occupancy 
between inner and outer shells.  Thus, 
when a Na$^+$ does enter a potassium-selective channel, the ion binds in
a configuration that does mimic the bulk hydration structure. In 
this case, however, that binding leads to a trapped ion that blocks
permeation.
For the well-known K-channel blockers, \Sr\ and \Ba,  
solution hydration
structures appear similar, as expected for ions of identical charge and
similar size. Six waters saturate the principal maxima of the RDFs,
though up to eight waters might be considered in broad settings.

Structural data for \Ba\  in K-channels was published
recently \cite{Jiang:2014}, though similar  data are unavailable for the
\Sr\ ion. In that \Ba\ crystal structure, eight oxygen atoms from the
innermost (S4) K-channel selectivity filter constitute the binding
contacts (Figure~\ref{fig:Hydmim_ion}). With the inner-shell occupancy 
(8) and
cavity radius (2.8\AA), the protein nicely mimics the hydration 
environment of \Ba.  Structural similarity thus 
supports the
hydration mimicry idea, but evidently the \Ba\  ion is too stably bound
(Figure~\ref{fig:Hydmim}).
\FloatBarrier

To summarize these local structural comparisons, hydration mimicry
appears applicable to all ions considered here, \emph{except} for K$^+$
in the potassium ion channels that inspired the concept.  In that case,
two more O atoms contact K$^+$ in channel binding sites than in aqueous
solution.  In all other cases for which crystal structures are
available, local solvation structures in crystallized channel binding
sites arrange to fit each ion with contact distances and numbers
anticipated from accurate solution information.
In some cases, hydration mimicry leads to
rapid ion permeation, but in other cases, mimicry
leads to trapped ions and block of permeation.



        

        
\section{CONCLUSIONS}

Hydration mimicry is a long-standing idea for 
functional design of membrane ion channels \cite{MacKinnon:2004el}.
Those concepts nevertheless enter
with coarse descriptors of ions and their interactions, \emph{i.e.},
sizes of protein binding sites  and of ions, of dielectric 
response of the environment, and of ion binding 
free energies  
described with those coarse factors. The work described here
builds the rigorous foundation for several of those factors, 
including ion sizes and hydration structures (Figure~\ref{fig:exp_QCT}, \ref{fig:gr}).  We emphasize that these primitive 
data, and specifically the neighborship analyses featured 
in Figure~\ref{fig:gr}, were not assured at the initiation of 
this research \cite{Rempe:Li,redbook,Mason:2015kw,Varma2006}.

Ion free energies (Figure~\ref{fig:Hydmim}) are natural assessments 
of those primitive descriptors and QCT \cite{redbook} 
addresses those free energies (Figure~\ref{fig:exp_QCT}), 
as it was designed to do (Figure~\ref{fig:diff_contri}).  
That theory provides a 
comprehensive description of the relative stability
of the end-points of the ion transfer process.
It also allows isolation of inner-shell 
(chemical contacts) and 
outer-shell interactions to illuminate mechanistic 
aspects of ion transfer. This information can be useful 
for designing membranes for specific phase transfer 
processes \cite{Cygan,Fu:2018}. 

For the membrane transport proteins assessed here, 
hydration mimicry applies in most, but not all, cases. 
Hydration mimicry applies to ions that permeate rapidly,
as well as ions that block ion permeation.
Nevertheless, the dynamics of the ion transfer process 
may also depend on the atomic-scale dynamical 
flexibility of the thermal systems considered \cite{Asthagiri:2010}.
       
\newpage
\begin{summary}[SUMMARY POINTS]
\begin{enumerate}
\item Properties that differentiate ions include size, charge, and coordination number.
\item Neighborship analyses describe
the structure of the $n$ closest
ligands to an ion, define \emph{direct} contacts, and 
reveal `\emph{split-shell}' coordination.
\item Direct-contact structure determinations
and QCT work together to seek mechanistic understanding 
of ion binding.
\item Interactions of metal ions with near-neighbors  are 
as strong as chemical interactions but selectivity
of ion transport depends on the balance of ion coordination
equilibria, including the aqueous solution endpoints of 
the transport.
\item The hydration mimicry concept applies to ions that 
permeate rapidly and to ions that block permeation; thus, hydration
mimicry does not necessarily guarantee rapid ion permeation.
\end{enumerate}
\end{summary}
\begin{issues}[FUTURE ISSUES]
\begin{enumerate}
\item Characterize the stability of protein binding sites 
by manipulation of water activity. What are 
probable $\mathrm{Ba}\left(\mathrm{H}_2\mathrm{O} \right)_m
\left(\mathrm{Thr} \right)_{n-1}{}^{2+}$  coordination cases 
for given Thr (threonine) solution concentrations?
\item For paradigmatic KcsA channel, are structural differences
between K$^+$ binding in the selectivity filter 
and in bulk water a key to K$^+$/Na$^+$ selectivity?
\end{enumerate}
\end{issues}

\section{ACKNOWLEDGMENT}
We thank Thomas L. Beck  for helpful discussions. Sandia National
Laboratories is a multi-mission laboratory managed and operated by
National Technology and Engineering Solutions of Sandia, LLC., a wholly
owned subsidiary of Honeywell International, Inc., for the U.S. DOE's
NNSA under contract DE-NA-0003525. This work was supported by Sandia's
LDRD program and performed, in part, at the Center for Integrated
NanoTechnology (CINT). This paper describes objective technical results
and analysis. Any subjective views or opinions that might be expressed
in the paper do not necessarily represent the views of the U.S. DOE or
the U.S. Government.

\newpage

\clearpage


\providecommand{\latin}[1]{#1}
\makeatletter
\providecommand{\doi}
  {\begingroup\let\do\@makeother\dospecials
  \catcode`\{=1 \catcode`\}=2 \doi@aux}
\providecommand{\doi@aux}[1]{\endgroup\texttt{#1}}
\makeatother
\providecommand*\mcitethebibliography{\thebibliography}
\csname @ifundefined\endcsname{endmcitethebibliography}
  {\let\endmcitethebibliography\endthebibliography}{}


\end{document}


\setcounter{page}{1} 


\maketitle


\section{Background: Supporting Information}

\begin{figure}[h]
\includegraphics[width=3.0in]{TOC_graphics.png}
\caption{Schematic illustrating the local structural focus of the
hydration mimicry idea and the utility of experimental structural data
for ion binding sites in protein pores (left) and molecular simulations
for ion hydration structures (right) to test the mimicry idea.
}\label{fig:TOC_graphics}
\end{figure}

We re-display our \emph{Table of Contents} graphic to support 
the discussion of the concept biomolecular mimicry on the basis of 
the local ion hydration environment
(Figure~\ref{fig:TOC_graphics}).
Here, ``\emph{local}'' structure refers to atoms that interact directly
with an ion. That local structural similarity may then provide a free
energy for ion binding that approximately equals the free energy for ion
hydration in bulk liquid water, leading to rapid ion permeation
(Figure~\ref{fig:Hydmim}). 

Applying the same idea to non-native ions offers explanations for
alternative transport behaviors (Figure~\ref{fig:Hydmim}). Without the
advantage of binding sites that mimic local hydration structure,
non-native ions may encounter large free energy barriers that lead to
rejection from the protein pore.  For ions similar in size to the
permeant ion, similarity in hydration structure is also anticipated,
leading to the same rapid ion permeation when pore binding sites arrange
like  local waters of ion hydration.  An exception is expected for an
ion of the same size that carries more charge than the permeant ion. 
Then binding of the more highly charged ion should block passage of the
permeant ion. Block ostensibly arises by trapping of the blocking ion
due to a more favorable free energy in the protein than experienced by
the ion in bulk aqueous
solution\cite{Jiang:2000,Jiang:2014,Piasta:2011bu,ye}.   

Here, we test the hydration mimicry concept by comparing local ion
hydration structure in liquid water, computed by molecular simulations,
to local solvation structure in crystallographically resolved channel
binding sites (Figure~\ref{fig:TOC_graphics}).  To relate the structures
to transport function, we compute ion transfer free energies between
liquid water and the same binding sites (Figure~\ref{fig:Hydmim}). 
Although the emphasis of the hydration mimicry idea is local, we also
take into account the influence of the surrounding environment on both
local structure and transfer free energies.  In some cases, we present
data from  earlier
studies,\cite{Varma2007,chaudhari2018SrBa} and in other
cases we compute new data.  

\section{Quasi-chemical theory: Supporting Information}

Note that, since Eq.~(3) is
balanced with respect to \emph{charge}, the $K_n$ do not 
involve the potential of the
phase.

\begin{figure}
    \includegraphics[width=3in]{QCT_Lambda_definition02.pdf}
    \caption{Shape of ion-water complex in PCM calculations, 
    showing change in dielectric boundary with change in the assigned 
    radius of the ion, $R_{\mathrm{M}^+}$, along with variation of water coordination 
    number, $n$. Results of outer-shell contributions to ion hydration free
    energy are minimally sensitive to $R_{\mathrm{M}^+}$ when the inner-shell solvation
    environment is filled with ligands, as illustrated in the bottom row.}
    \label{fig:QCT_lmd_def}
\end{figure}

\subsection{Potential of the phase: Supporting Information}
The conclusion of the main text 
\begin{eqnarray}
		2  q e \Delta \Phi = -\Delta 
\left\lbrack\mu_{\mathrm{M}^{q+}}^{(\mathrm{ex})}-\mu_{\mathrm{X}^{q^-}}^{(\mathrm{ex})}\right\rbrack ~,
\label{eq:surfpot}
\end{eqnarray}
is strikingly simple,
so it is worthwhile to proceed further with some of the details. 
Matching the single-ion chemical potentials in two
coexising phases then focuses on the
junction potential 
\begin{eqnarray}
q e \Delta  \Phi =   -  \Delta \mu_{\mathrm{M}^{q+}} ^{(\mathrm{ex})} 
- kT  \Delta 
\ln\rho_{\mathrm{M}^{q+}}  ~.
\label{eq:electrochempot2}
\end{eqnarray}
Without ionic effects, the leftside of Eq.~\eqref{eq:electrochempot2}
vanishes because $q=0$.  With ionic effects and the $-q$ ionic 
counter-ion, we combine to find
\begin{multline}
2 q e \Delta  \Phi =   -  \Delta \left\lbrack
 \mu_{\mathrm{M}^{q+}} ^{(\mathrm{ex})}  - \mu_{\mathrm{X}^{q^-}} ^{(\mathrm{ex})} \right\rbrack \\
- kT \Delta 
 \ln\left\lbrack \frac{\rho_{\mathrm{M}^{q+}} }{\rho_{\mathrm{X}^{q^-}}}\right\rbrack ~.
\label{eq:electrochempot3}
\end{multline}
Since the bulk compositions are electroneutral, the rightmost term
vanishes. The physical conclusion is that $\Delta  \Phi$ achieves
neutral compositions even though chemical-scale effects are 
not necessarily simply balanced.  

The argument leading to Eq.~\eqref{eq:electrochempot2} shows how to
treat more general cases of ionic composition.  For example, a mixture
of $q-q$ electrolytes MX and MY, brings us to
\begin{multline}
3 q e \Delta  \Phi =   -  
\Delta \left\lbrack
\mu_{\mathrm{M}{}^{q+}} ^{(\mathrm{ex})}  
    - \mu_{\mathrm{Y}{}^{q-}} ^{(\mathrm{ex})}  
    - \mu_{\mathrm{X}^{q^-}} ^{(\mathrm{ex})} 
\right\rbrack \\
- kT \Delta 
\ln\left\lbrack 
 \frac{
 \left(\rho_{\mathrm{X}{}^{q-}} + \rho_{\mathrm{Y}{}^{q-}}\right)
 }{
\left(\rho_{\mathrm{X}{}^{q+}}\right)\left( \rho_{\mathrm{Y}{}^{q+}}\right) }\right\rbrack ~,
\label{eq:electrochempot4}
\end{multline}
having used the neutrality $\rho_{\mathrm{M}{}^{q^+}} =
\rho_{\mathrm{X}{}^{q-}} + \rho_{\mathrm{Y}{}^{q-}}$ in each
phase. If one of the electrolytes is infinitely dilute, then
Eq.~\eqref{eq:surfpot} is recovered after noting the condition of
transfer equilibrium for the dilute component
Eq.~\eqref{eq:electrochempot2}. Generally, the additional  composition
factors of Eq.~\eqref{eq:electrochempot4} remain.

\section{Ion hydration free energy: Supporting Information}\label{sec:hydra_fenergy}

Considering the several ingredients that
are combined to evaluate the net free energy, we base the present 
discussion on the examples of Rb$^+$ and Ca$^{2+}$
(Figure~\ref{fig:diff_contri}).  Those free energy contributions
(Figure~\ref{fig:diff_contri}) are all substantial on a chemical energy
scale; 
that is, comparable to traditional chemical bonds. 
Free energy contributions for the other divalent ions
considered here (\Mg,~\Sr,~\Ba) appear in the Supplementary Information (Figure S3).

Noting the trends in
each free energy contribution (Figure~\ref{fig:diff_contri}), 
the gas phase association term becomes more favorable
with increased $n$ for Rb$^+$.  
The contribution
to hydration free energy from waters beyond the inner-shell clusters becomes less favorable and gradually approaches zero
as cluster size increases.  That outer-shell
contribution also varies with boundaries, explored here for $\lambda$ set between
the first maximum and minimum of the RDF.  The variation with boundary becomes small upon reaching clusters
with full occupancy of the inner shell, at $n$=4 for Rb$^+$ (vertical gray shading in Figure~\ref{fig:diff_contri}).  
Further description of the free energy contributions for \Ca hydration follows.

The combination in Eq.~\eqref{eq:1} is correct for any $\lambda$ and $n$, but
comparison with the specific simplification arriving at Eq.~\eqref{eq:4}
shows that the rightside of Eq.~\eqref{eq:4} should track $-kT \ln
p(n)$, as it depends on $n$, provided that other aspects of the theory
are satisfactory.   The results for \Ca (Figure~\ref{fig:diff_contri}) do
support that view.  Then for chemically reasonable $\lambda$, the net
free energy is insensitive to $n$ for values of $n$ associated with a
fully occupied inner hydration shell (Figure~\ref{fig:gr}). 

The variation of the free energy results of Figure~\ref{fig:diff_contri}
with $\lambda$ and $n$ is encouragingly simple.  Nevertheless, our
experience has been that the dielectric model for the outer-shell
contributions, here PCM, can be problematic when the ion inner-shell
occupancy differs substantially from full occupancy.\cite{Sabo:2013gs} Figure~\ref{fig:QCT_lmd_def} illustrates this issue.
Contrapositively, the thermodynamic free energies can be satisfactory
when this inner-shell occupancy is saturated because of the
balance between the cluster and ligand terms of the rightmost
contribution of Eq.~\eqref{eq:4}.  Then, sensitivity to adjustment of
boundaries for dielectric models, which are not molecularly realistic,
is moderated because the adjusted boundaries are somewhat buried in the
cluster, though otherwise balanced.   This point emphasizes
that the highlighted  rightmost (cluster) contribution of
Eq.~\eqref{eq:4} is the principal theoretical challenge for QCT free
energies, particularly after acknowledging that the $-kT \ln p(n)$
contribution, dropped in Eq.~\eqref{eq:4}, might be helpfully
estimated by widely accessible simulation calculations, as
in some prior work.\cite{sabo:h2,Chaudhari:2014wb}

This example (Figure~\ref{fig:diff_contri}) used the PCM
approach\cite{Tomasi:2005tc} to implement Eq.~\eqref{eq:4}.  We assumed
without comment that the value of $\lambda$ used for the clustering
analysis with statistical mechanical theory Eq.~\eqref{eq:4} can be used
also as a dielectric boundary for the Ca$^{2+}$ ion. Consequences
(Figure~\ref{fig:diff_contri}) are reasonable agreement with a tabulated
thermodynamic free energy, reasonable description of $\ln p(n)$, and
that the free energy predictions are encouragingly insensitive to
$\lambda$. The last of these points is significant since continuum
models are typically sensitive to the boundaries,\cite{Ashbaugh:2008hm}
and those boundaries are not independently defined by physical
principle, indeed\cite{Linder:2004eu} are ``\ldots a complicated
function of density, temperature, and molecular parameters \ldots''

Nevertheless, we can turn this argument around.  Assigning a radius
$R_{\mathrm{Ca}^{2+}}$ for the central ion, then ascertaining a QCT free
energy prediction (Eq.~\eqref{eq:4}) insensitive to
$R_{\mathrm{Ca}^{2+}}$, we ask for the  dielectric radius of a single
sphere for the  Ca$^{2+}$ ion to match that free energy within the Born
model.\cite{rashin1985reevaluation}  This approach thus provides
rational estimates of  dielectric radii.

In closing this section, we note again that evaluation of $K^{(0)}_{n}$
is accessible with widely available tools of computational chemistry.
Therefore, QCT addresses realistic distribution of charge for the
clusters formed in Eq.~\eqref{eq:2charged}. Realistic distribution of
charge might include charge transfer, general inductive adjustment of
charge, thus effects that might be captured by classical molecular
polarizibility models. Those electronic structure calculations can
include van~der~Waals interactions, as a general matter.  Though our
discussions here have centered on atomic ions, the issues of
distributions of charge within ions remains; those issues are likely to
acquire further specificity for molecular ions.

\section{Ion transfer free energies: Supporting Information}

Transfer free energies reported here
quantify the difference in absolute solvation free energies between an ion in a 
protein binding site relative to the ion in aqueous solution. 
To assess the hydration mimicry idea,
we present new data for \Mg\ and \Ca\ binding to Mg-selective and 
Ca-selective binding sites,
and summarize prior studies of
K$^+$, Na$^+$, \Sr, and \Ba~binding to
two distinct K-channel binding sites, an
interior site and the innermost site (S4).

\begin{table}
\begin{tabular}{| c | c | c| c|}
\hline    
Ion       & PCM       & \multicolumn{2}{c|} {$\mu_{\mathrm{M}^{2+}}^{(\mathrm{ex})}$} \\
 &  &  \multicolumn{2}{c|} {(kcal/mol)} \\
 &  $\epsilon$  & QCT & EXP\cite{Marcus:1994ci} \\
\hline
			& 2   & -351.7 &  \\
     			& 4   & -396.6 &  \\
Mg$\left(\mathrm{H}_2\mathrm{O}\right)_6{}^{2+}$ & 10  & -441.6 & \\ 
     			& 30  & -424.6 &  \\
     			& 50  & -441.6 &  \\
     			& 80  & -443.0 &  -441.2 \\
     \hline
			& 2   & -278.3 &  \\
     			& 4   & -318.5 &  \\ 
 Ca$\left(\mathrm{H}_2\mathrm{O}\right)_8{}^{2+}$ & 10  & -345.5 & \\
     			& 30  & -358.1 &  \\
     			& 50  & -360.5 &  \\
     			& 80  & -362.2  & -363.5  \\
\hline
\end{tabular}
\caption{Effect of PCM $\epsilon$ on solvation 
free energy
for \Mg\ and \Ca\ with $n=6$ and 8 waters, respectively, relevant to  
Figure~\ref{fig:Hydmim_ion}.}\label{table:1}
\end{table}

Transfer free energies (Table~\ref{table:1}) and 
structural comparisons fully 
support hydration
mimicry as an essential mechanism for \Mg~and \Ca~transport 
through Mg-transporter or
Ca-channels. The inner-shell structures at the binding site, 
Mg(H$_2$O)$_6{}^{2+}$ and Ca(H$_2$O)$_8{}^{2+}$, are identical to
the most stable hydration structures in bulk liquid water environments 
(Figure~\ref{fig:Hydmim_ion}). Each protein mimics the electrostatic 
environment of
the more distant hydration environment, represented by high dielectric values
in a PCM model, to compensate for the missing waters of hydration 
experienced by ions at the protein binding sites (Table~\ref{table:1}).  In other words,
hydration mimicry occurs non-locally, in contrast to the traditional vision (Figure~\ref{fig:Hydmim}).

 Transfer free energies for the monovalent K$^+$ and Na$^+$  (Figure~\ref{fig:wat_chan_ene_V4}) reflect the permeation, rejection, and trapping behaviors of Figure~\ref{fig:Hydmim}, and are consistent with experimental observations. The permeation profile occurs for transfer of K$^+$ from water, where four (4) waters define the peak
 in RDF at a distance of 2.7~\AA, to an interior binding site represented by four (4) diglycine molecules (GG) that ligate the ion with eight (8) oxygen atoms at a distance of 2.8~\AA. Rapid permeation by K$^+$ and selective ion binding of K$^+$ over the smaller Na$^+$ ion is attributed to crowding in flexible binding sites with high density of ligands, relative to water. Crowding requires constraints on ligand number for Na$^+$ (denoted by open symbol) and preferentially destabilizes the smaller ion. Trapping of Na$^+$ occurs upon release of constraints on ligand number, when the binding site distorts and becomes uncrowded with only five (5) ligating oxygens located at a distance of 2.4~\AA~(denoted by an asterisk). Lack of hydration mimicry corresponds to Na$^+$ rejection and K$^+$ permeation profiles, while
 hydration mimicry corresponds to trapping of
 Na$^+$.

In the context of free energies for Sr$^{2+}$ and Ba$^{2+}$  transfer
between water and K$^+$ channel binding sites, 
the slightly smaller Sr$^{2+}$ requires approximately an additional 30 kcal/mol
to take the ion out of the hydration environment (Figure~\ref{fig:exp_QCT}). The
additional stability comes mainly from a stronger association free
energy for Sr$^{2+}$ in water relative to Ba$^{2+}$, as expected due to
the smaller size and higher charge density of Sr$^{2+}$. This additional
free energy must be compensated by the channel for both ions to have
similar binding affinities. 

Similar to the hydration results,  Sr$^{2+}$ is 30 kcal/mol more
favorably solvated in the S4 binding site than 
Ba$^{2+}$ (Figure~\ref{fig:wat_chan_ene_V4}). 
Again, the additional stability comes from a stronger association free
energy for the smaller Sr$^{2+}$ in the S4 binding site.  Similar
outer-shell contributions are expected for the two ions in a given
environment, as observed (Figure~\ref{fig:gr}), due to the 
similarity in cluster
size, as well as identical ligand chemistry and number of ligands.   

Comparing ion binding to the S4 binding site relative to water
($\mu^{\mathrm{(ex)}}_{\mathrm{X} \mathrm{-S4}}$ versus
$\mu^{\mathrm{(ex)}}_{\mathrm{X}}$), the calculated binding free
energies show only minor differences between values of the same ion 
(Figure~\ref{fig:wat_chan_ene_V4}). Interestingly, components of the
free energy show substantial differences.  The association free energy
for both Sr$^{2+}$ and Ba$^{2+}$ in the S4 binding site of threonine
oxygens is 30~kcal/mol more favorable than in water due to the stronger
ion-ligand interactions from threonine oxygens.  In contrast, the
outer-shell contributions are about 30~kcal/mol less favorable in S4
due to the larger size of the ion-ligand clusters in the S4 binding
site. These opposing trends lead to a transfer free energy of approximately 
for both ions, suggesting that the blocking site environment $\it{can}$
provide an additional 30 kcal/mol  to stabilize the Sr$^{2+}$ ion. 

\begin{figure}[hbt]
\centering
    \begin{subfigure}[b]{0.43\textwidth}
    \includegraphics[width=1.45in]{Mg_channel.png}\hfill
		\includegraphics[width=1.45in]{Mg_7.png}
		\end{subfigure}
		\vskip\baselineskip
		\begin{subfigure}[b]{0.43\textwidth}
		\includegraphics[width=1.45in]{Ca_channel.png}\hfill
		\includegraphics[width=1.45in]{Ca_8.png}
		\end{subfigure}
		\vskip\baselineskip
		\begin{subfigure}[b]{0.43\textwidth}
		\includegraphics[width=1.45in]{Ba_KcsA.png}\hfill
		\includegraphics[width=1.45in]{Ba_8.png}
		\end{subfigure}
		\vskip\baselineskip
		\begin{subfigure}[b]{0.43\textwidth}
		\includegraphics[width=1.45in]{K_S2_V3.png}\hfill
		\includegraphics[width=1.45in]{K_W4.png}
		\end{subfigure}
		\vskip\baselineskip
		\begin{subfigure}[b]{0.43\textwidth}
		\includegraphics[width=1.45in]{Na_Gly_4.png}\hfill
		\includegraphics[width=1.45in]{Na_W4.png}
		\end{subfigure}
		\vskip\baselineskip
        \caption{Comparison of solution optimized ion-water clusters 
        (right) with 
        X-ray crystal structures of ions
		inside the channels (PDB, left top to bottom,
		4U9L\cite{Payandeh2013}, 4MS2\cite{Tang2014}, 
		4PDR\cite{Jiang:2014}, 1K4C\cite{Zhou:2001vo}, and 2ITC\cite{Lockless:2007ex}). Except for the K$^+$ case, the 
		ion coordination inside the 
		channel corresponds to the local ion hydration structure, 
		consistent with the  
		hydration mimicry idea.}\label{fig:Hydmim_ion}
	\end{figure} 



\section{Acknowledgment}
We thank Thomas L. Beck for helpful discussions.
Sandia National Laboratories is a multi-mission laboratory managed and
operated by National Technology and Engineering Solutions of Sandia,
LLC., a wholly owned subsidiary of Honeywell International, Inc., for
the U.S. Department of Energy’s National Nuclear Security Administration
under contract DE-NA-0003525. This work was supported by Sandia's LDRD
program. This paper describes objective technical results and analysis.
Any subjective views or opinions that might be expressed in the paper do
not necessarily represent the views of the U.S. Department of Energy or
the United States Government.

\clearpage

\bibliographystyle{biochem}
\bibliography{baBib,QCTbib,MypaperBib,Ca_Mg_rdf_expt,Sr_bib,Ca_Mg,Additional_refs}
